\documentclass[aps,pre,twocolumn,amsmath,amssymb,showpacs,showkeys,floatfix]{revtex4-1}
\usepackage{epsfig}
\usepackage{graphicx}
\usepackage{bm}
\usepackage{amsmath}
\usepackage{physics}
\usepackage{xcolor}
\usepackage{amsmath, nccmath, mathtools}
\usepackage{hyperref}

\begin{document} 

\title{Chaos and localized phases in a two-body linear kicked rotor system}
\author{Anjali Nambudiripad}
\thanks{These two authors contributed equally.\\}
\author{J. Bharathi Kannan}
\thanks{These two authors contributed equally.\\}
\author{M. S. Santhanam}
\affiliation{Indian Institute of Science Education and Research, Pune 411 008, India.}
\date{\today}

\begin{abstract}
Despite the periodic kicks, a linear kicked rotor (LKR) is an integrable and exactly solvable model in which the kinetic energy term is linear in momentum. It was recently shown that spatially interacting LKRs are also integrable, and results in dynamical localization in the corresponding quantum regime. Similar localized phases exist in other non-integrable models such as the coupled relativistic kicked rotors. This work, using a two-body LKR, demonstrates two main results; firstly, it is shown that chaos can be induced in the integrable linear kicked rotor through interactions between the momenta of rotors. An analytical estimate of its Lyapunov exponent is obtained. Secondly, the quantum dynamics of this chaotic model, upon variation of kicking and interaction strengths, is shown to exhibit a variety of phases -- classically induced localization, dynamical localization, subdiffusive and diffusive phases. We point out the signatures of these phases from the perspective of entanglement production in this system. By defining an effective Hilbert space dimension, the entanglement growth rate can be understood using appropriate random matrix averages.
\end{abstract}

\maketitle

\section{Introduction}
The kicked rotor is a paradigmatic model of Hamiltonian chaos and continues to remain relevant in the context of current interest in the effects of interactions in chaotic systems \cite{Chirikov-1971,SanPauBha22}. Physically, a single-kicked rotor system can be thought of as a pendulum to which energy is externally imparted through periodic delta kicks. For sufficiently strong kick strengths, the classical system displays chaotic dynamics and results in unbounded diffusive growth of mean energy. 

In the corresponding quantum regime, a different behaviour emerges -- the initial diffusive growth of mean energy is sustained only for a timescale of $t_b \sim O(1/\hbar^2)$. Beyond this timescale, diffusive growth is strongly suppressed by destructive quantum interference effects resulting in localization of wavefunction. In a seminal work, Fishman {\it et al.} showed that the quantum kicked rotor (QKR), at large kick strengths, could be mapped to an Anderson-type model with a quasi-disordered potential \cite{FisGrePra82}. Unlike Anderson localization in disordered systems, dynamical localization is unrelated to genuine disorder or intrinsic randomness and is a consequence of deterministic but chaotic system dynamics. 

In last two decades, QKR has emerged as a popular model for understanding the physics of localization through theory and experiments; see Ref. \cite{SanPauBha22} for a recent review of QKR and its variants. Dynamical localization in QKR has been experimentally observed through cold atoms periodically kicked by optical lattices formed by a counter-propagating standing wave of light \cite{MooRobBha94}. Quasi-periodic kicked rotor models with three incommensurate frequencies have allowed for precise observation of the metal-insulator Anderson transition in $d \geq 3$ \cite{LemChaSzr09}.


Presently, there is significant interest in the interplay between classical chaos and quantum localization in the presence of interactions \cite{TohHuiMcc2022,CaoRosMas2022,LelRanAda2020,NotIemRos18,NotSilFaz2020}. One of the questions being vigorously investigated is the fate of dynamical localization under the effect of interactions. Since localization is a delicate quantum effect, it was generally believed that it might not survive under the effect of many-body interactions. However, it is now well appreciated that, as was argued first in the case of weakly interacting electrons, many-body localization could be preserved under suitable conditions \cite{BasAleAlt06,OgaHus07,PalHus10,BarPolMoo12,IyeOgaRef2013}. It is then reasonable to expect that many-body variants of dynamical localization should also exist.  When many QKRs interact nonlinearly with one another, a localized phase is shown to exist for a finite number of rotors \cite{NotIemRos18,NotSilFaz2020}, and in the  interacting kicked Lieb-Liniger system \cite{ChiRan21,VuaRan21, RylRozGal20,VuaRan23}.  Apart from a localized phase, a subdiffusive phase appears in the border between the dynamically localized and delocalized phases \cite{NotSilFaz2020}, though localization vanishes in the thermodynamic limit. However, there exist some counter-examples in which dynamical localization survives even in the thermodynamic limit.

One such counter-example is the {\it linear} kicked rotor (LKR) system (Hamiltonian is linear in momentum, see Eq. \ref{eqn:1}). Introduced by Berry in 1984 \cite{Ber84}, LKR is an integrable model. Curiously, an interacting system of linear kicked rotors is also integrable. Localization in momentum representation exists (for particular choice of parameters) and exact integrals of motion were obtained \cite{KesGanRef16}. This is a case of dynamical many-body localization (DMBL). A related example is the interacting relativistic kicked rotors (LKR appears in its high energy limit) \cite{RozGal17}, a non-integrable but non-chaotic model and displays dynamical localization. Although localization emerges in these models, corresponding classical phase space is either entirely regular (as in LKR) or a mixture of mostly regular layer with a small chaotic component (as in the relativistic model). Hence, the observed quantum localization largely arises due to regular components in classical phase space. This still leaves the question of whether DMBL can survive and new dynamical phases can emerge if LKR is modified such that it displays predominantly chaotic dynamics.

This question assumes importance in the light of two recent experimental results with periodically kicked interacting cloud of atoms \cite{TohHuiMcc2022,CaoRosMas2022}, in a setup that allows for tunable interactions. They have demonstrated that even though interactions did not sustain localization, a full-blown diffusive phase is absent. Dynamical localization is replaced by a quantum subdiffusive (weakly diffusing) phase even when the underlying classical dynamics remain chaotic. Thus, it is of interest to study if LKR shows these novel emergent phases.

Following this discussion, several questions arise in the context of LKR. Since an $N$-body interacting linear kicked rotor is an integrable model (provided the interactions are smooth functions), in some sense, it would lack the dynamical richness of a non-integrable system; there would neither be mixed phase space nor chaos. Hence, firstly, how to induce non-integrability and chaos in this model. Secondly, if chaos is induced in the interacting LKR model, does the localized phase survive ? Do any new dynamical phases appear in this system. 

To answer these questions about dynamical phases, we consider an interacting (two-body) linear kicked rotor model and demonstrate that (a) chaos can be induced in interacting LKR if the momenta are coupled, and this leads to qualitatively rich dynamical features, (b) upon variation of parameters, a variety of phases -- classically induced localization, dynamical localization, subdiffusive and diffusive phases -- exist in LKR. 
Interactions in kicked rotor models can be introduced in several ways; usually by coupling the rotors in position space and less commonly in momentum space \cite{AdaTodIke88, HirTakMas95, WeiMulPic95, She00, WenLei09, BonKudMon10, SanArn20}. In this work, the momenta of the rotors are coupled together, generating classical chaotic dynamics in LKR. Analytic estimates of Lyapunov exponents are obtained to indicate the presence of chaos. In contrast, coupling the position variables does not generate chaotic dynamics for any interaction strength \cite{KesGanRef16}. It might be pointed out that considerable amount of work had focussed on modeling interactions between rotors using noise \cite{She83,OttAntHan84,Coh91,Coh_2_91,Coh_3_91,ShiHuB95,Bor96,SchHenLut08}, dissipation \cite{Coh_3_91,DitGra90,DitGra_2_90,DyrMil95}, and non-linearity \cite{BenCasPik91,She93,PikShe08,Mie04,GarShe09,FlaKriSko09,MulAhnPik09,Gli11}, though they do not study LKR system. \\


 Section \ref{sec:LKR} briefly introduces the LKR model, followed by a detailed discussion on the classical dynamics of momentum coupled linear kicked rotor ($m$-LKR) in Section  \ref{sec:mLKR}. The quantum dynamics and corresponding dynamical phases in the parameter space of $m$-LKR are discussed in section \ref {sec:Loc_mLKR}. This is followed by the study on entanglement production in $m$-LKR in section \ref{sec:EE}. Section \ref {sec:conclusions} presents a brief summary of the results and conclusions.

\section{Linear Kicked Rotor}
\label{sec:LKR}
As pointed out earlier, Fishman {\it et. al.} showed that the Floquet operator of QKR can be mapped to the dynamics of a single particle tight-binding lattice with quasi-periodic potential \cite{FisGrePra82}. However, the lattice model does not have an analytical solution. In order to solve the mapping analytically and obtain an accurate description of the lattice model, a modification to the KR model was introduced \cite{Ber84,PraGreFis84,GreFisPra82,FisGrePra84,FisGrePra82} in the form of linear kicked rotor. As this name suggests, LKR has a linear, instead of quadratic, dependence on the momenta. The tight-binding lattice model corresponding to LKR is also called the Maryland model \cite{Bar85,Wat92}. \\

The Hamiltonian of the single-particle linear kicked rotor (LKR) model is given by
\begin{equation}
    H_{\rm LKR}(t) = 2\pi{\alpha}p+ K\cos(x)\sum_{n=-{\infty}}^{\infty}\delta(t - nT),
    \label{eqn:1}
\end{equation}
where $0 \leq x < 2\pi$ and $-\infty < p < \infty$, and the classical dynamics takes place on cylindrical phase space. Physically, it can be thought of as a particle on a ring moving at constant speed, completing $\alpha$ number of rotations in a time period $T$ of kicks. These periodic delta-function kicks have a kicking strength $K$. Due to the existence of a complete set of integrals of motion (IOM), energy stays localized when $\alpha$ is irrational. When the resonance condition is met, the IOMs break down leading to delocalization in momentum space \cite{Ber84}.

For $H_{\rm LKR}(t)$ in Eq. \ref {eqn:1} with cosine kicking potential, the resonance condition is met when $\alpha \in \mathbb{Z}$. Both rational and irrational values of $\alpha$ give rise to IOMs in momentum space leading to localization in the momentum variable. For $\alpha \in \mathbb{Z}$, these IOMs diverge leading to resonant (quadratic) mean energy evolution. On the other hand, if the kicking potential is a smooth function that could be Fourier expanded in terms of $\cos(x)$, then irrational values of $\alpha$ lead to IOMs and localization. Further, rational $\alpha$ satisfies the resonant condition leading to ballistic transport in the system \cite{Ber84}.

Throughout this work, $\alpha$ is chosen to be an irrational number since we are interested in the effect of interactions on localization. It ensures the system is far from the resonance conditions, even in the presence of interactions.  The time evolution of mean kinetic energy, rather than the total energy, will be used as an indicator of dynamical localization \cite{KesGanRef16,RozGal17}. The classical trajectories generated by Eq. \ref {eqn:1} with irrational $\alpha$ are regular finite loops around its cylindrical phase space, and energy evolves periodically with time, implying that $\langle p^2 \rangle \approx $ constant. Thus, the classical LKR model of Eq. \ref {eqn:1} is devoid of chaos at all times for any value of $K$. In this case, one observes a clear correspondence between the classical and quantum dynamics of the system.

In general, the classical or quantum transport in the kicked rotor systems can be conveniently quantified based on the typical time-dependence of the mean squared angular momenta given by
\begin{equation}
    \left\langle p^{2}\right\rangle \sim t^{\beta},
    \label{Eqn:transp_1}
\end{equation}
where $\beta$ is the exponent. If quantum dynamics is considered, then $\langle . \rangle$ represents the usual expectation value over the time-evolving state. In the classical regime, $\langle . \rangle$ represents an average taken over an ensemble of initial conditions.
If the dynamics is localized in momentum space, then $\beta = 0$. If $0<\beta<1$, it corresponds to sub-diffusion, and $\beta=1$ corresponds to linear diffusion. When $1<\beta <2 $, we will have anomalous diffusion, while the particular case of $\beta=2$ corresponds to ballistic transport. Ballistic transport is observed in LKR when the resonance condition is met \cite{KesGanRef16}. 

\subsection{Spatially coupled LKR}
To construct an interacting model with $N$ rotors, a straightforward choice is to couple the position variables of the rotors. One possibility is to generate a system with the Hamiltonian given by
\begin{align}
H(t) = ~ & 2\pi \sum_i^N \alpha_i p_i + \sum_{i \ne j} H_{\rm int}(x_i - x_j) ~ + \nonumber \\
     & K \sum_i V(x_i) \sum_{n=-\infty}^{\infty} \delta(t-nT),
\label{Hx-model}
\end{align}
where $V(x_i)$ is the potential of $i$-th rotor and $H_{\rm int}(x_i - x_j)$ represents rotor-rotor interactions that is some function of $x_i - x_j$. If $V(x_i)$ and $H_{\rm int}(x_i - x_j)$ are chosen to be any smooth functions of position coordinates that can be expanded in a Fourier representation, then it was shown\cite{KesGanRef16} that IOMs will always exist for any $K$ and irrational $\alpha_i$. Consequently, it would not display chaotic dynamics. In the corresponding quantum regime, we anticipate that localization would be preserved as the absence of classical non-integrability has a strong influence \cite{RozGal17,KesGanRef16}. This type of dynamical many-body localization (DMBL) arises due to the presence of additional IOMs bounded in momentum space \cite{KesGanRef16}. The non-interacting LKR and the spatially interacting model in Eq. \ref {Hx-model} are both integrable, and hence a correspondence between the classical and quantum regimes can be expected to exist.

\section{Momentum-coupled linear kicked rotor}
\label{sec:mLKR}
The previous works on spatially interacting LKR \cite{KesGanRef16,RozGal17} argued that the emergence of additional IOMs and the DMBL are universal and would survive even if the model was made non-integrable through some generalization. In this work, the validity of this claim is tested by introducing interactions which couple momenta of the linear kicked rotors. A prime motivation for introducing coupling in momentum variable is to induce and probe chaotic dynamics in an otherwise regular system. Recall that any smooth coupling through position variables will not induce chaotic dynamics. Then, the Hamiltonian for the $N$-particle momentum coupled LKR ($m$-LKR) is
\begin{equation}
    {H} = {H_0}(\mathbf{p}) + H_{\rm int}(\mathbf{p})+ {V}(\mathbf{x})\sum_{n=-\infty}^{\infty}\delta(t-nT),
    \label{eqn:Ham_2}
\end{equation}
where ${H_0}(\mathbf{p})=2\pi\sum_i \alpha_i p_i$ is the unperturbed part which has a linear dependence on momenta, $H_{\rm int}(\mathbf{p})$ is a nonlinear perturbation that encodes the interaction in momenta between the particles, and ${V}\mathbf{(x)}$ is the potential modulated by a series of kicks. In this paper, $m$-LKR is studied for the case of two rotors ($N=2$), but the formalism implemented is general and can be extended to any $N$. For $N=2$, explicit forms for the terms in the Hamiltonian of Eq. \ref {eqn:Ham_2} is 
\begin{equation}
\begin{aligned}
H_0(p_1,p_2) & = 2\pi\alpha_{1}{p_{1}}+2\pi\alpha_{2}{p_{2}},\\
H_{\rm int}(p_1,p_2) & =  k_{p} ~ p_{1} ~ p_{2}, \\
V(x_{1},x_{2}) & = K \left[ \cos({x_{1}})+\cos({x_{2}}) \right],
\end{aligned}
\end{equation}
where $\alpha_1 \ne \alpha_2$  are irrational numbers and $k_p$ is the interaction strength. The kicking potential $V(x_1,x_2)$ of strength $K$ acts on the rotors only at integer times, {\it i.e.,} with a periodicity of $T=1$. Though the form of interaction $H_{\rm int}$ was specifically chosen for ease of computation, it can be generalized with the provision that it needs to be a smooth function to make physical sense. 


\subsection{Classical dynamics of the $m$-LKR}
\label{sec:Class_LKR}
The classical dynamics of $m$-LKR is studied by simulating the stroboscopic map (from just after $t$-th kick to just before $(t+1)$-th kick) and is given by
\begin{equation}
\begin{gathered}
p_{1}^{t+1}=p_{1}^{t}+K \sin \left(x_{1}^{t}\right), \\
p_{2}^{t+1}=p_{2}^{t}+K \sin \left(x_{2}^{t}\right), \\
x_{1}^{t+1} =\left(x_{1}^{t}+2 \pi \alpha_{1}+ k_{p}p_{2}^{t+1} \right)~ \rm{mod} ~2\pi, \\
x_{2}^{t+1} =\left(x_{2}^{t}+2 \pi \alpha_{2}+ k_{p}p_{1}^{t+1} \right)~ \rm{mod} ~2\pi. 
\end{gathered}
\label{Eqn:class_map}
\end{equation}
In this, the superscript $t$ or $t+1$ represents the integer times at which kicks are imparted to the system, and the numbering of rotors is given in the subscripts. If $k_p=0$, then the system is integrable for any $K \ge 0$. As we shall show below, for a fixed $K$ and $k_p \gg 1$, the system becomes non-integrable and eventually nearly fully chaotic. Further, if we multiply the first two equations by
$k_p$, and denote scaled momenta as 
\begin{align}
P_1^{t+1}=k_p p_1^{t+1} \;\;\;\; \mbox{and} \;\;\;\; P_2^{t+1}=k_p p_2^{t+1},
\label{eqn:scaled_p}
\end{align} 
then the nature of classical dynamics is determined by a single scaled parameter $K_s=k_p K$. The average classical energy of the rotors (averaged over an ensemble of initial conditions) is $\langle E\rangle_{c} = \langle p^{2}_{1}+p^{2}_{2}\rangle$. In the chaotic limit, this can be easily estimated within the quasi-linear approximation to be $\langle E\rangle_{c} = D_0 t$, where $D_0=K^2$ is the diffusion coefficient. It must be noted that $k_p$ should be sufficiently large for chaos to set in. However, once this limit is reached, diffusion rate $D_0$ (within the quasi-linear approximation) does not depend on $k_p$.

\subsection{Chaotic dynamics of the $m$-LKR}
\begin{figure}[h!]
	\includegraphics[width=0.46\linewidth]{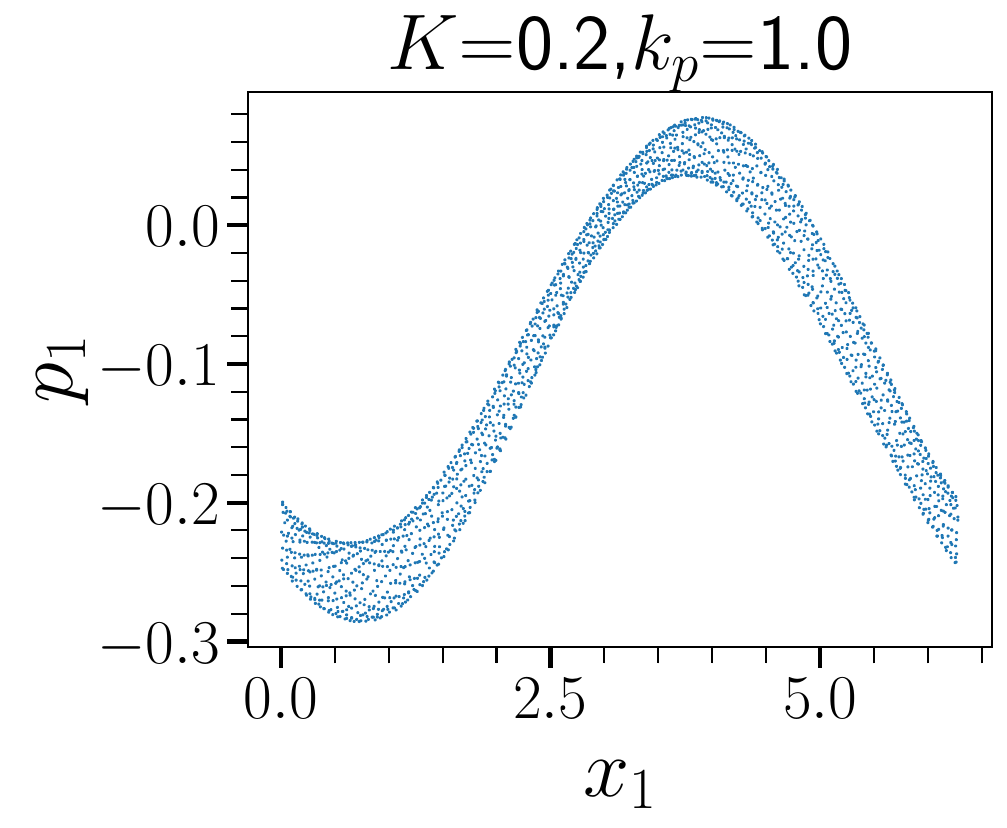}
	\includegraphics[width=0.48\linewidth]{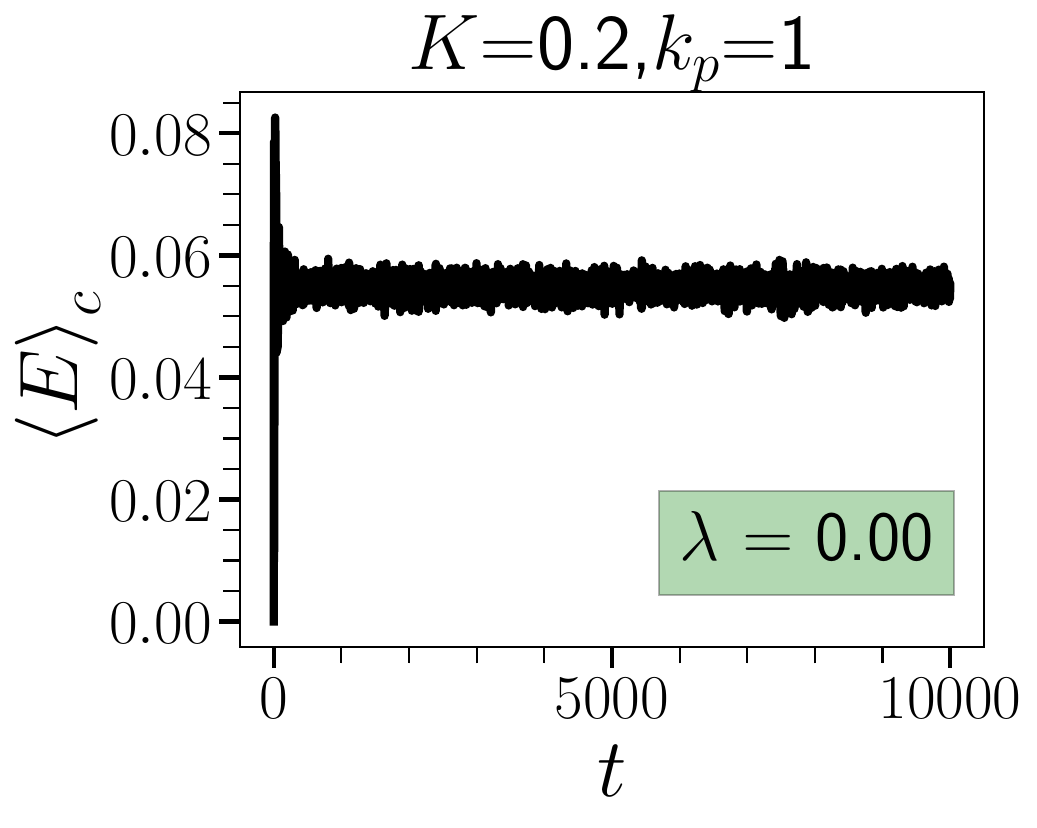}

	\includegraphics[width=0.46\linewidth]{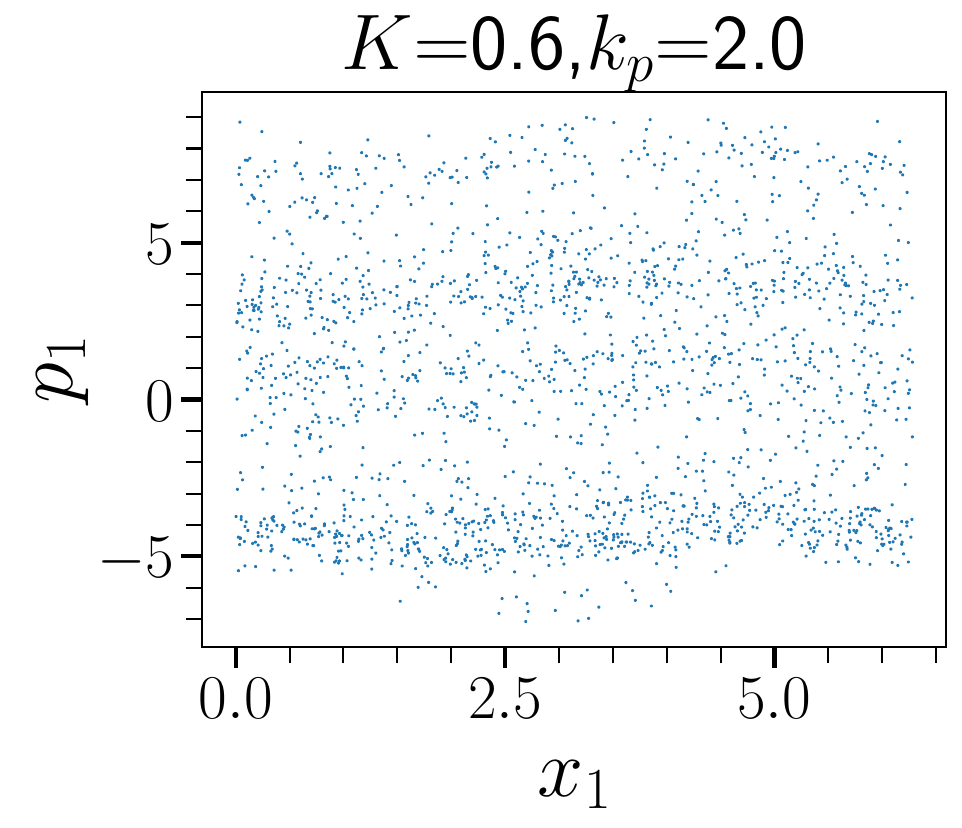}	
	\includegraphics[width=0.48\linewidth]{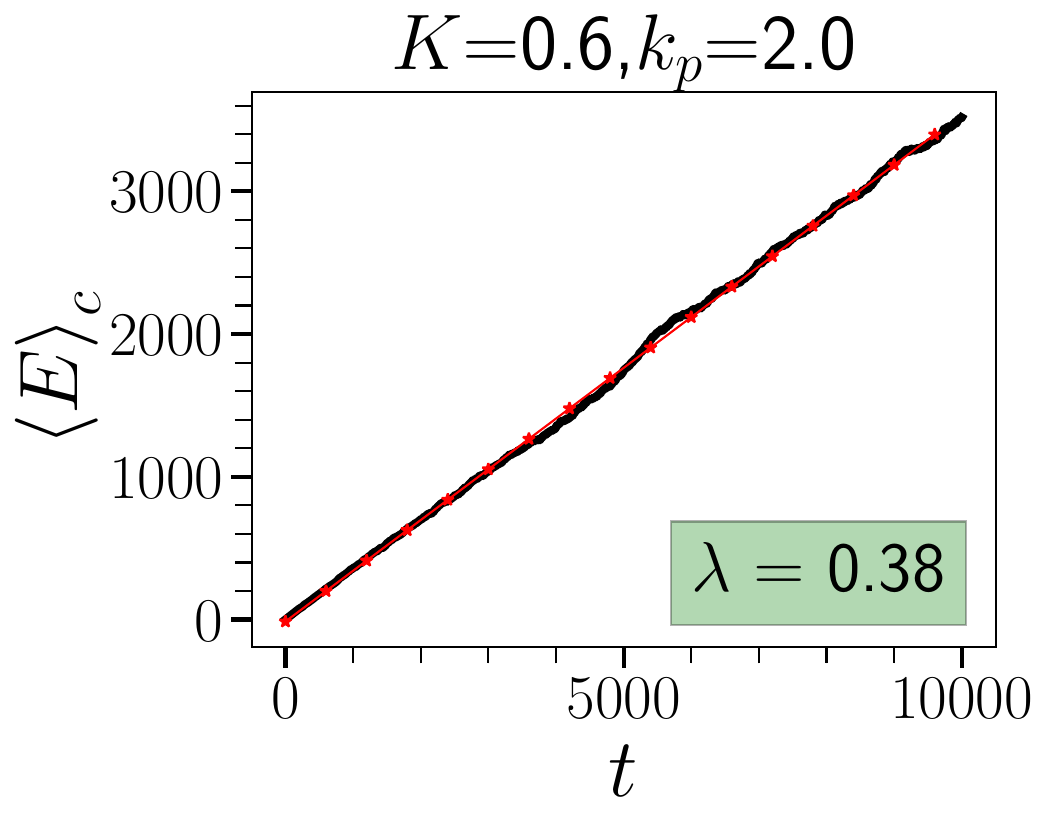}

	\includegraphics[width=0.46\linewidth]{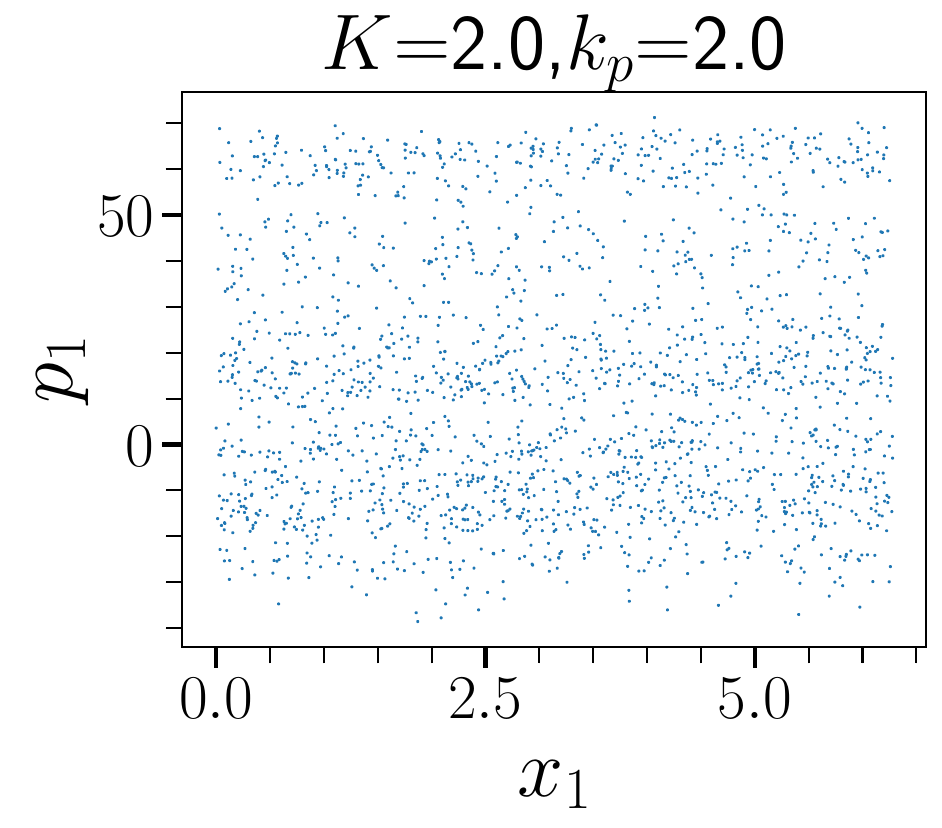}	
	\includegraphics[width=0.48\linewidth]{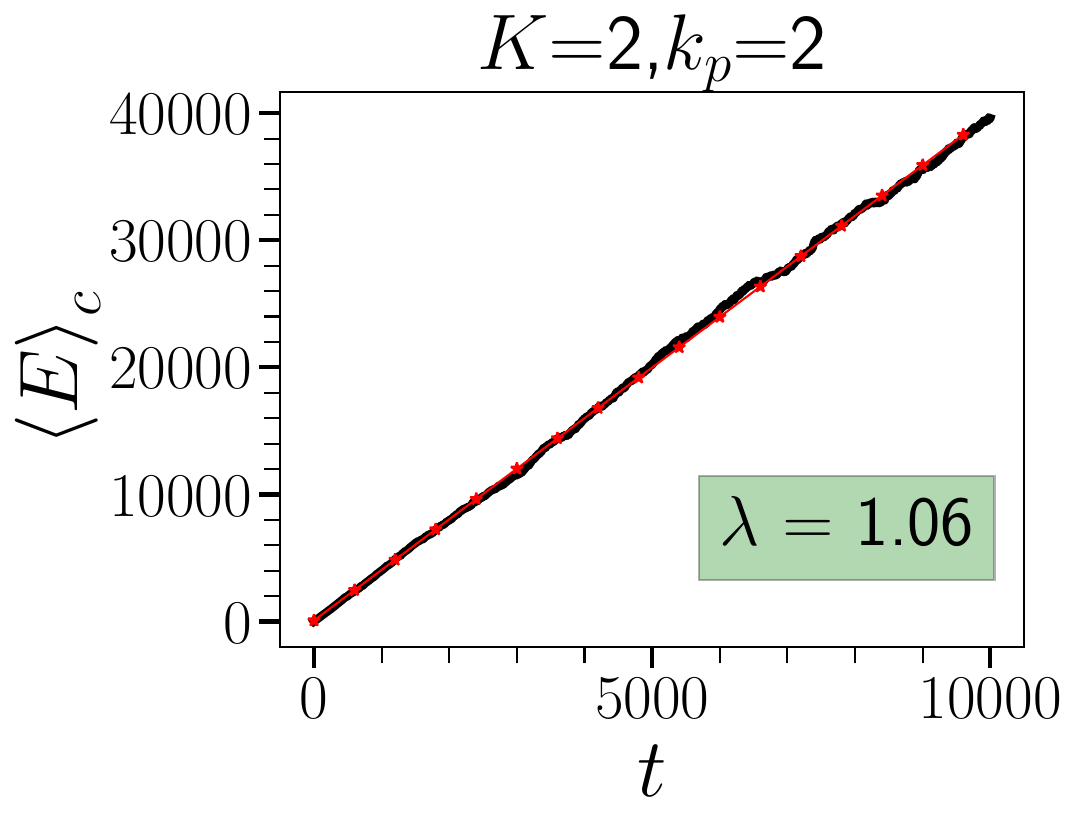}

	\includegraphics[width=0.46\linewidth]{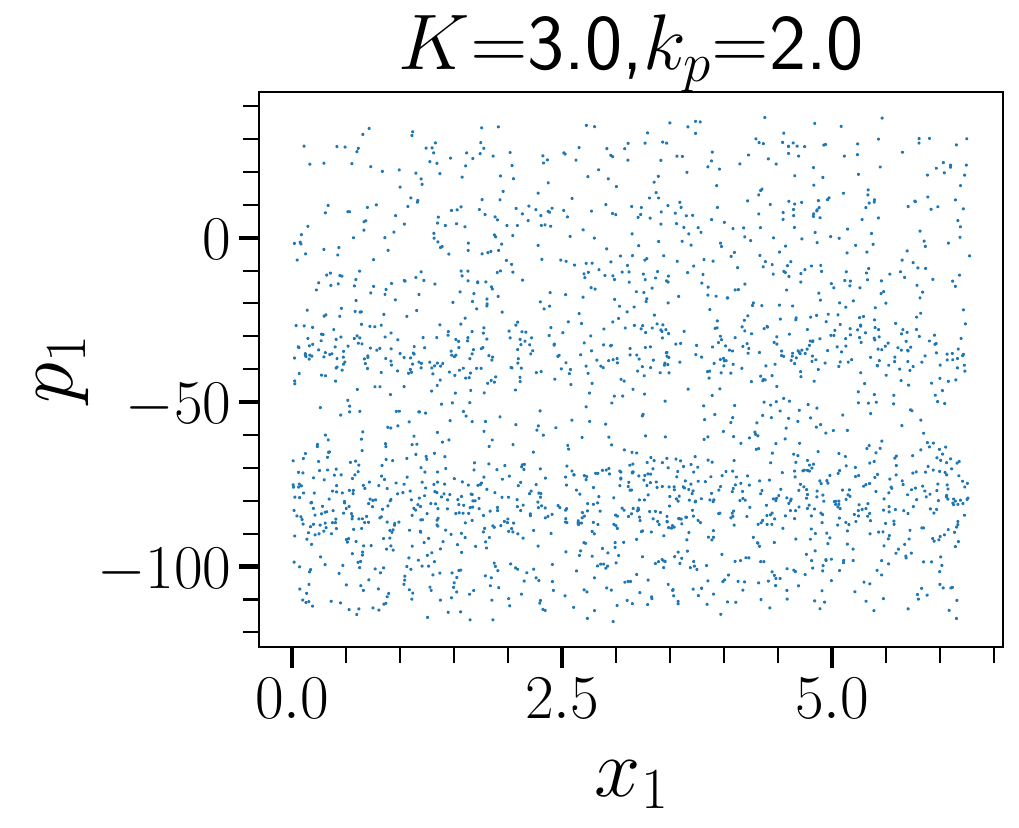}	
	\includegraphics[width=0.48\linewidth]{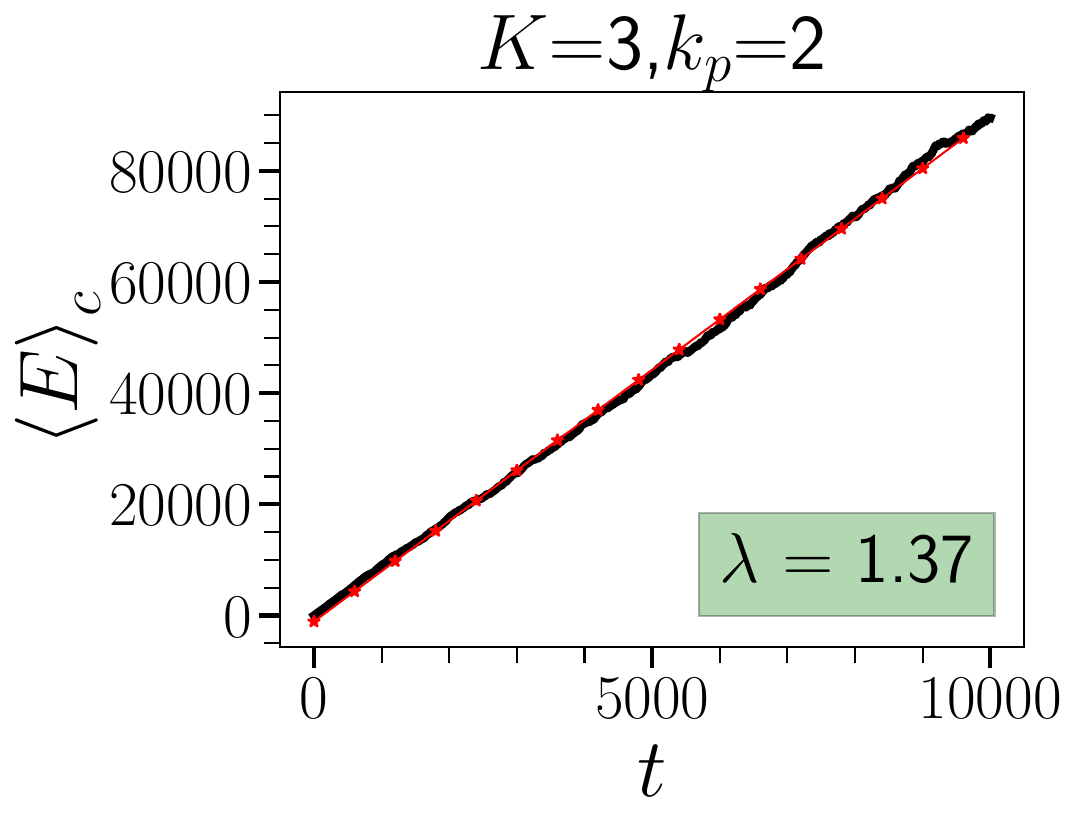}

	\includegraphics[width=0.46\linewidth]{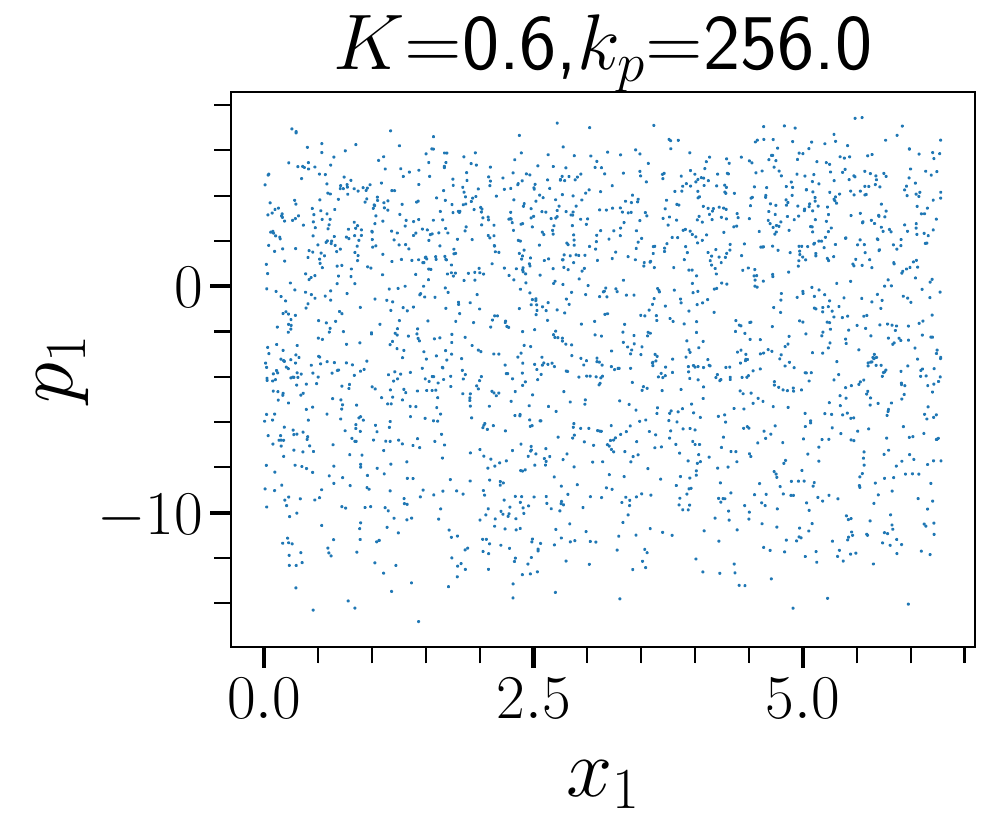}	
	\includegraphics[width=0.48\linewidth]{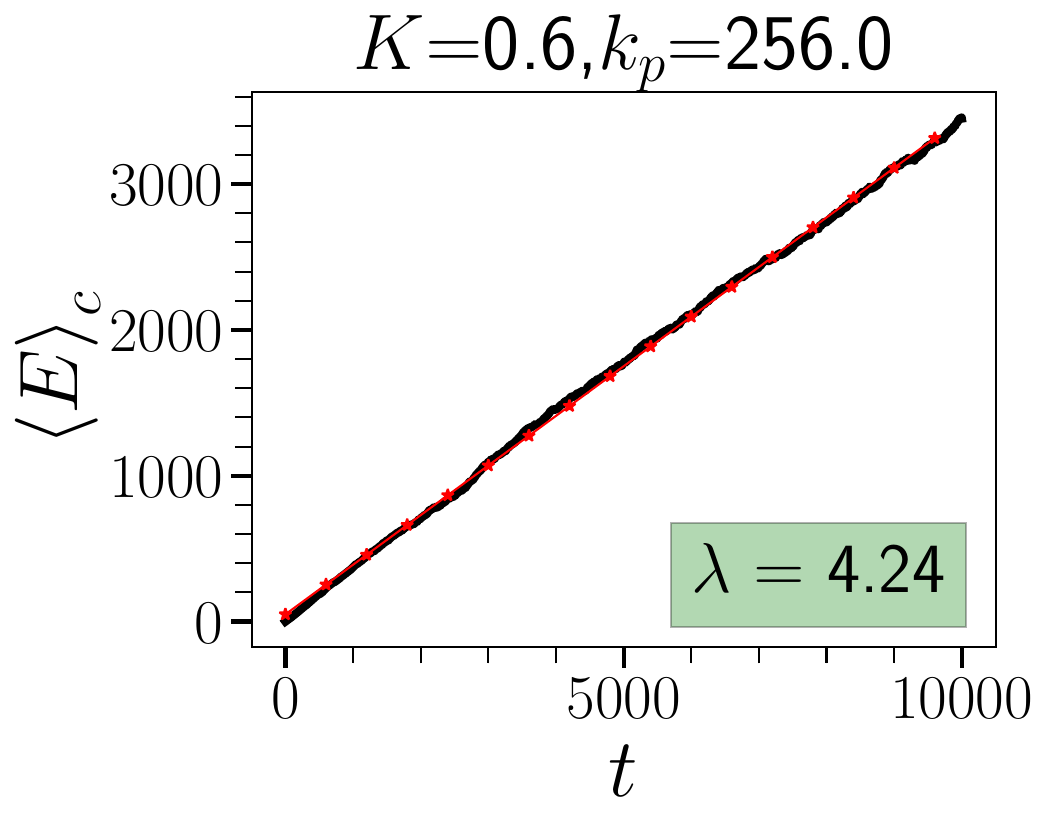}
    \begin{picture}(0,0)
	\put(-206,462){(a)}
    \put(-86,462){(b)}
    \put(-206,360){(c)}
	\put(-86,355){(d)}
	\put(-206,264){(e)}
    \put(-86,257){(f)}
    \put(-206,169){(g)}
    \put(-86,163){(h)}
    \put(-206,76){(i)}
    \put(-86,70){(j)}
	\end{picture}
\caption{Left column shows the $(x_1,p_1)$ projection of the stroboscopic section for a selection of parameters $(K,k_p)$. The right column shows the time evolution of mean classical energy $\langle E\rangle_{c} = \langle p_1^2 + p_2^2 \rangle$ (corresponding quantum energy growth is shown in Fig. \ref{fig:2}). The numerically computed Lyapunov exponent $\lambda$ is also shown. Other parameters are $\alpha_{1}=\sqrt{3}$, $\alpha_{2}=\sqrt{5}$. The system is evolved for $t=10^4$ kicks, and the classical energy is averaged over $10^4$ initial conditions. The red curve is the analytical estimate of classical energy growth obtained through quasi-linear approximation.} 
\label{fig:1}
\end{figure}

The classical mean energy $\langle E\rangle_{c}$ and stroboscopic map are obtained by numerically simulating Eq. (\ref {Eqn:class_map}), starting from an ensemble of initial conditions. The $\langle E\rangle_{c}$ is averaged over $10^4$ initial conditions randomly chosen from $\mathbf{x}\in[0,2\pi]$ and $\mathbf{p} = 0.0$. This choice corresponds to an initial wave function (a $\delta$-function in momentum space at $\mathbf{p} = 0.0$) used to simulate quantum dynamics. Throughout this paper, though only the $(x_1,p_1)$ section is shown, its counterpart, namely $(x_2,p_2)$ section, shows qualitatively similar dynamics as the former.  This system has 4 degrees of freedom and therefore no easy means of visualisation exist. Thus, the stroboscopic plots in Fig. \ref{fig:1}(a,c,g,e,i) display a projection of the dynamics on the $(x_1,p_1)$ plane. Then, we expect to see trajectories crossing one another which is an artifact of projection. Despite this drawback, it is still useful to broadly distinguish chaos from pre-dominantly regular dynamics, and it might not be useful in the mixed dynamics regimes. Figure \ref {fig:1}(a) shows the stroboscopic plot for $K=0.2$ and $k_p=1$ such that $K_s=0.2 < 1$. For this choice, we expect the system to be in a perturbative regime. As evident from the presence of invariant curves in Fig. \ref{fig:1}(a), the dynamics is mostly regular and large stochastic layers are absent. Then $\langle E\rangle_{c}$, shown in \ref {fig:1}(b), tends to a constant (barring small fluctuations). The numerical estimate of the Lyapunov exponent further supports this observation: $\lambda \approx 0.00$. As $K_s \to 0$, the coupled system does not exhibit chaotic dynamics.

In contrast, Fig. \ref{fig:1}(c) shows chaotic dynamics in the section for $K_s=1.2$ ($K=0.6$ and $k_p=2$) and no regular structures are visible to the resolution of our simulations. Consistent with chaos, $\langle E\rangle_{c}$ has a linear evolution with time, and $D_0 \approx 0.29$ (estimated from simulations) is reasonably close to the quasi-linear estimate of $K^2 = 0.36$. In this case, as we might anticipate, the largest Lyapunov exponent is a small positive number and gives $\lambda = 0.38$. As $K_s$ increases further, the sections shown in Figs. \ref{fig:1}(e,g,i) display chaotic dynamics for $K_s=4~ (K=2, k_p=2), K_s=6 ~(3,2)$ and $K_s=153.6~ (0.6, 256)$ respectively. The different colors in the sections correspond to different initial conditions. Since chaos has already ensued, the largest Lyapunov exponents (computed from simulations) are large positive numbers, namely $\lambda = 1.06, 1.37$ and $4.24$ respectively for $K_s=4, 6$ and $153.6$. For these parameters, the mean energy displayed in Figs. \ref {fig:1}(f,h,j) are all linear in time, and the values of $D_0$ (obtained from simulations), respectively, are 4, 9, and 0.36. These values have an excellent agreement with the quasi-linear estimate of $D_0 \approx K^2$.

Further, we obtain an approximate analytical estimate for the largest Lyapunov exponent in the limit of $K_s \gg 1$. To do this, note that the period-1 fixed points of the scaled version of the map in Eq. \ref {Eqn:class_map} is given by
\begin{align}
\left( x_1 = 2 l_1 \pi, ~ x_2 = 2 l_2 \pi, P_1 = -2\pi \alpha_2, P_2 = -2\pi \alpha_1 \right),
\label{inicond}
\end{align}
where $l_{1},l_{2} \in \mathbb{Z}$. By linearising the map about this fixed point or any other arbitrary point $\theta = (x_1, x_2, P_1, P_2)$ in phase space\cite{ott_2002}, we obtain the Jacobian matrix to be
\begin{align}
J(\theta) = \begin{pmatrix}
1 & K_s \cos x_2 & 0 & 1 \\
K_s \cos x_1 & 1 & 1 & 0 \\
K_s \cos x_1 & 0 & 1 & 0 \\
0 & K_s \cos x_2 & 0 & 1 
\end{pmatrix}.
\label{eq:jacob}
\end{align}
Note that, due to linear dependence on momentum in the Hamiltonian (see Eq. \ref{eqn:Ham_2} and \ref{eqn:scaled_p} ), the linearized dynamics is independent of $P_1$ and $P_2$. Then, the Lyapunov exponent evaluated at $\theta$ can be obtained, after $t$ kicks, from the eigenvalues of $[J(\theta)^{\rm T}]^t ~ [J(\theta)]^t$. Let $\sigma^2(\theta,t)$ represent its largest eigenvalue. Then, the largest Lyapunov exponent is 
$\lambda_{\rm max} = \langle \lim_{t \to \infty} (1/t)\log\sigma(\theta,t) \rangle$, where averaging is performed over all possible $\theta$. If $\lambda_{\rm max} > 0$, then it indicates chaotic dynamics. We obtain an approximate expression for $\lambda_{\rm max}$ as a function of $K_s$ by computing the eigenvalues of $[J(\theta)^{\rm T}]^t ~ [J(\theta)]^t$. This can be done as follows: if the operation $[J(\theta)^{\rm T}]^t ~ [J(\theta)]^t$ is performed analytically, and the limit $K_s \to \infty$ is taken, then after $t$ kicks, we obtain the dominant contribution to be

\begin{align}
\lim_{K_s \to \infty} [J(\theta)^{\rm T}]^t ~ [J(\theta)]^t \propto K_s^{2t}  
\begin{pmatrix}
a_1 & 0 & 0 & 0 \\
0 & a_2 & 0 & 0 \\
0 & 0 & 0 & 0 \\
0 & 0 & 0 & 0
\end{pmatrix}.
\label{eq:appx_jacob}
\end{align}

\begin{figure}[t]
	\includegraphics[width=\linewidth]{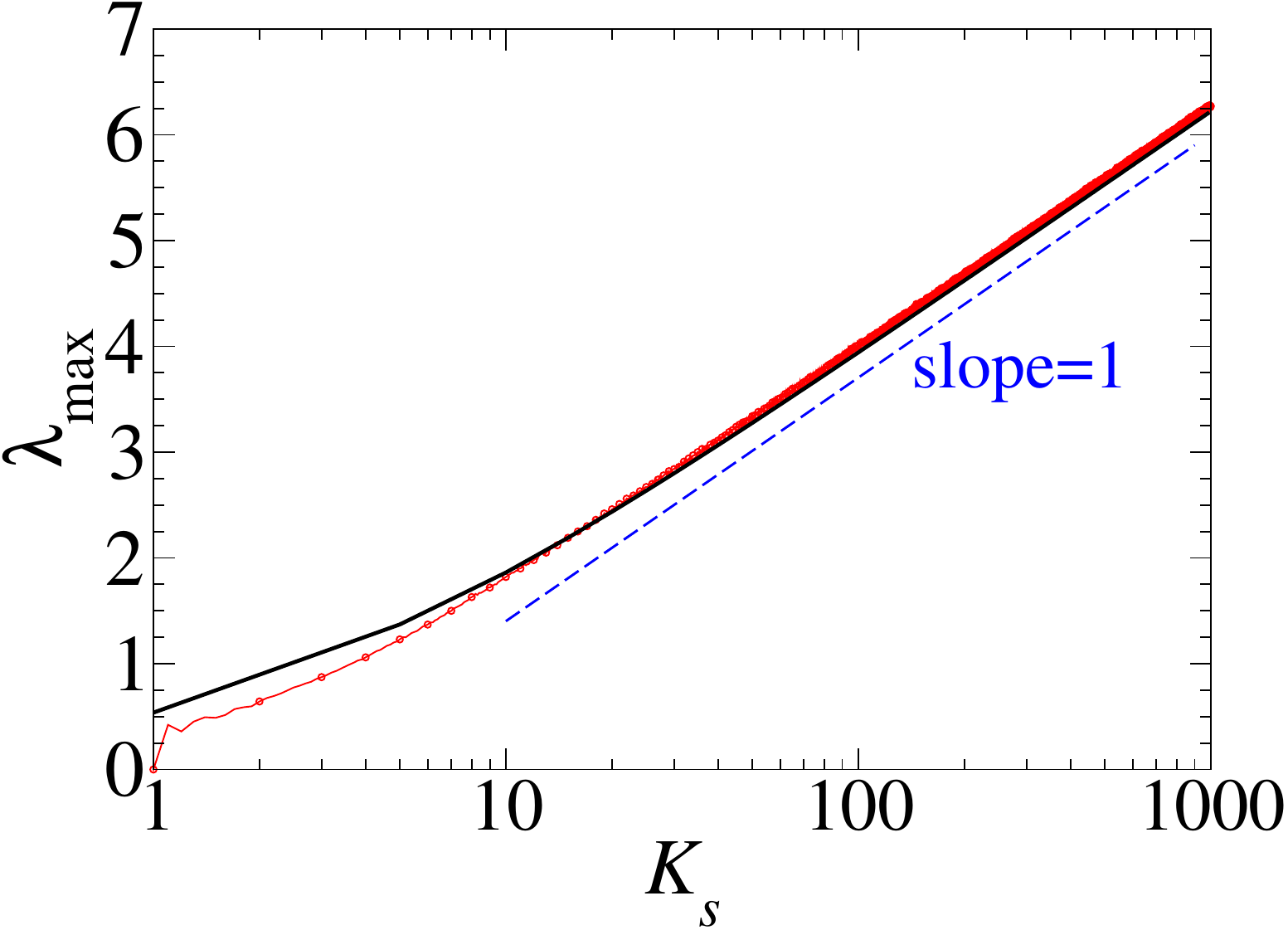}
\caption{Largest Lyapunov exponent $\lambda_{\rm max}$ shown as a function of $K_s$ in semi-log axis. Red symbols are computed from the map (\ref {Eqn:class_map}), and the black line is obtained by evolving using the Jacobian in Eq. \ref {eq:jacob}. For $K_s \gg 1$, a good agreement is seen between the largest Lyapunov exponent computed by both methods. The dashed (blue) line has slope 1 and is a guide to the eye. Each value of the Lyapunov exponent is averaged with respect to $\theta$.}
\label{fig:lyaexp}
\end{figure}

In this approximation, at $t$-th kick, the highest power of $K_s$ is $2t$, and all the lower powers of $K_s$ have been ignored, and $0 \le a_1,a_2 \le 1$ are constants (independent of $K_s$) that arise from averaging over $\theta$. Using this, the dependence of the largest eigenvalue on $K_s$ becomes
\begin{align}
\sigma^2 \approx K_s^{2t}.
\label{largeev}
\end{align}
From this, we can infer that $\lambda_{\rm max} \approx \ln(K_s)$. This was checked against a direct computation of $\lambda_{\rm max}$ by evolving the tangent vectors of the map in Eq. \ref{Eqn:class_map}. Figure \ref{fig:lyaexp} shows $\lambda_{\rm max}$ as a function of $\ln(K_s)$ obtained using these two different methods. The red symbols represent $\lambda_{\rm max}$ directly obtained from the map, and the black line indicates that obtained by evolving $[J(\theta)^{\rm T}]^t ~ [J(\theta)]^t$ for $t = 8000$ kicks. The slope being unity in the semilog plot suggests that $\lambda_{\rm max} \sim \ln(K_s)$. We also observe a good agreement between these two curves especially for $K_s \gg 1$. This approximation breaks down for $K_s \approx 1$ and in fact over-estimates the dependence since we ignored the lower order terms in Eq.\ref {eq:appx_jacob}. We might note that the logarithmic dependence of $\lambda_{\rm max}$ on $K_s$ is reminiscent of similar log-dependence, at large chaos parameters, in the case of other popular classical kicked systems, namely, kicked rotor \cite{SanPauBha22} and the kicked top \cite{Haa10}.


\subsection{Quantum dynamics of $m$-LKR}
The quantum dynamics of interacting systems, whose classical limits possess integrable to chaotic transitions, continues to attract attention. However, most of the studies deal with individual subsystems that are chaotic, but in the case of $m$-LKR, there is no classical chaos in the subsystems in the absence of interactions. In fact, as discussed in Sec. \ref{sec:Class_LKR}, the entire system becomes chaotic only if the rotors interact through their momenta. The Floquet operator for $m$-LKR in Eq. \ref{eqn:Ham_2} can be written down as
\begin{equation}
\begin{aligned}
   {\widehat{F}}= & \exp(-i\widehat{V}) ~ \exp({-i(\widehat{H}_{0}+\widehat{H}_{\rm int})(\textbf{p})}) \\
   = & \exp({-i\widehat{V}}) ~ \exp({-i\widehat{H}_{0}(\textbf{p})}) ~ \exp({-i\widehat{H}_{\rm int}(\textbf{p})}).
\end{aligned}
\label{Eqn:Floq_1}
\end{equation}
The last form arises because the interaction term satisfies the commutation relation $\left[H_{0}, H_{int}\right]=0$. As in the classical case, the parameters $\alpha_1$ and $\alpha_2$ are chosen to be irrational numbers. In the quantum model, by scaling the momenta as $P_{1} = \alpha_{1}p_1$ and $P_{2} = \alpha_{2}p_2$, the relevant scaled parameters become $\widetilde{k}_p = k_p/\alpha_1 \alpha_2$, $\widetilde{K}=K/T$ and scaled Planck's constant $\hbar_s = \hbar/T$. Throughout this work, the Planck's constant is fixed at $\hbar = 1$. As before, we also fix $\alpha_1=\sqrt{3}, \alpha_2=\sqrt{5}$, and $T=1$. 

\begin{figure}[t]
	\includegraphics[width=0.46\linewidth]{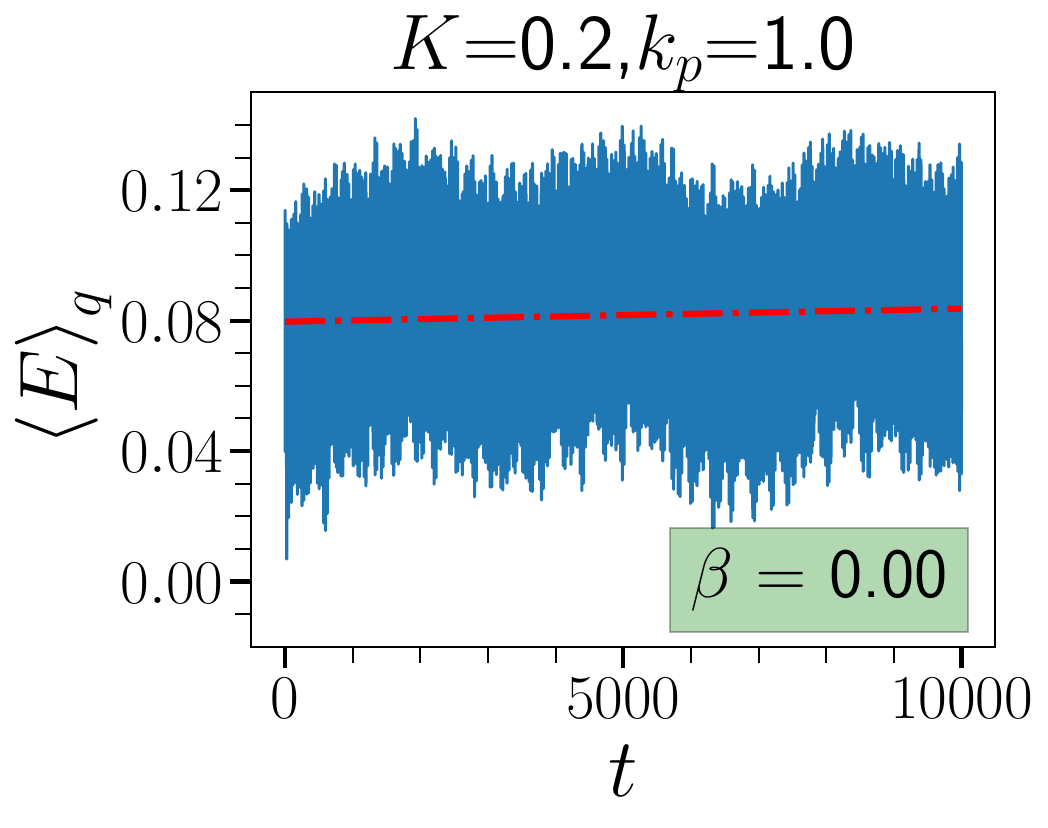}
    \includegraphics[width=0.46\linewidth]{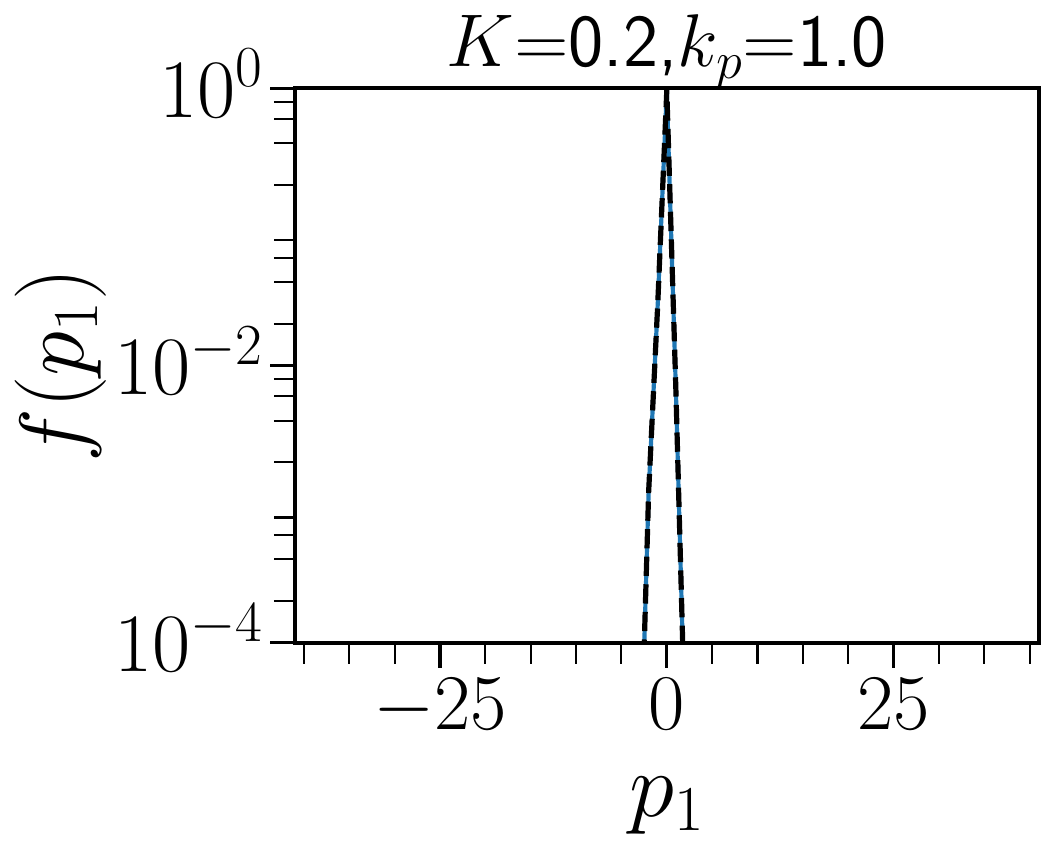}

	\includegraphics[width=0.46\linewidth]{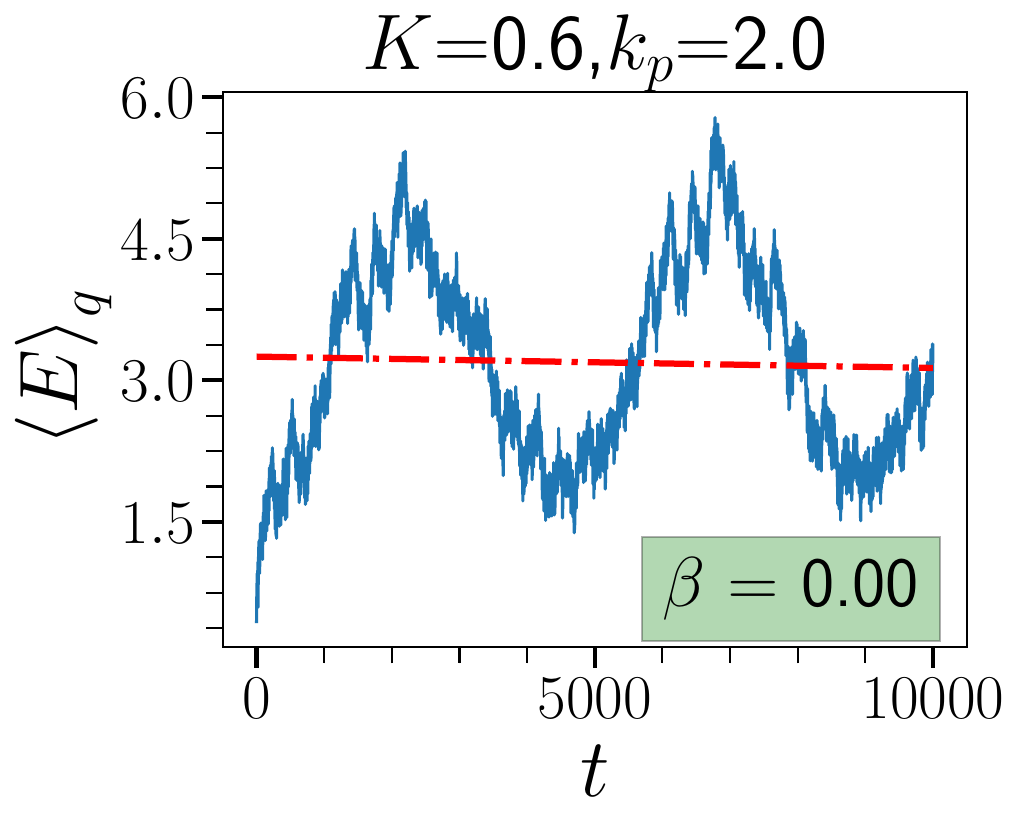}
    \includegraphics[width=0.46\linewidth]{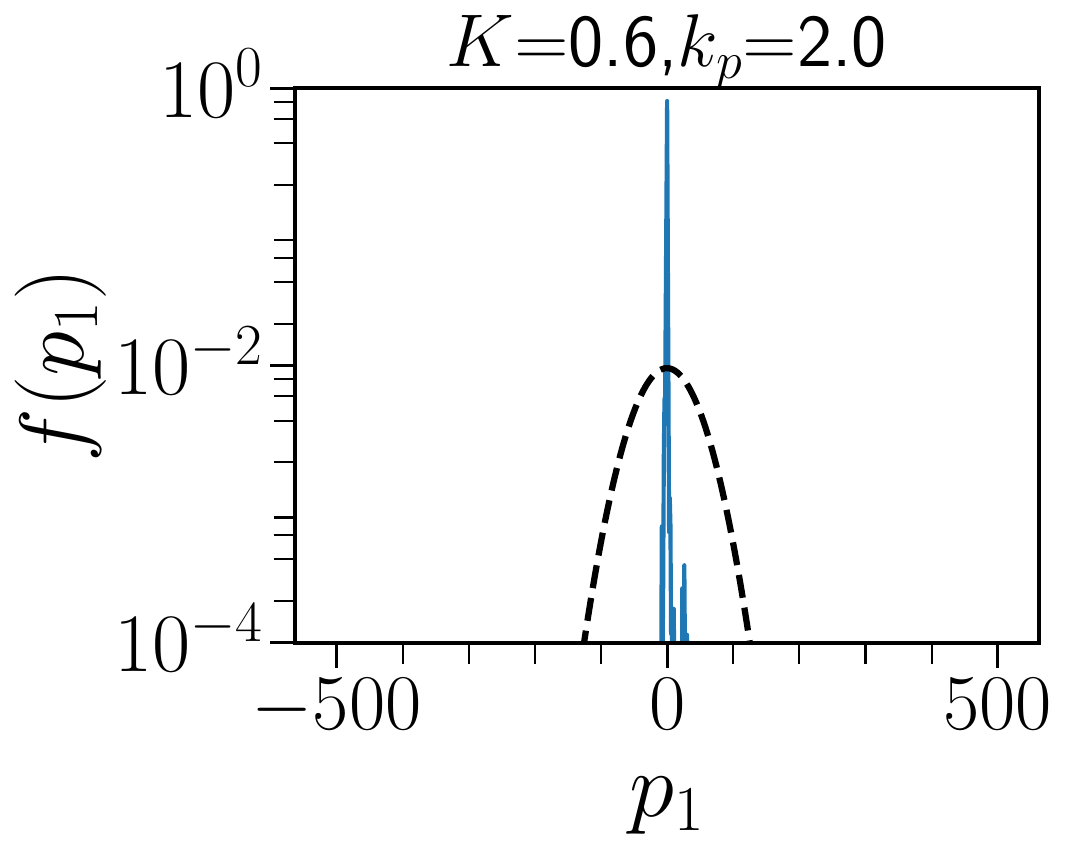}
    
	\includegraphics[width=0.46\linewidth]{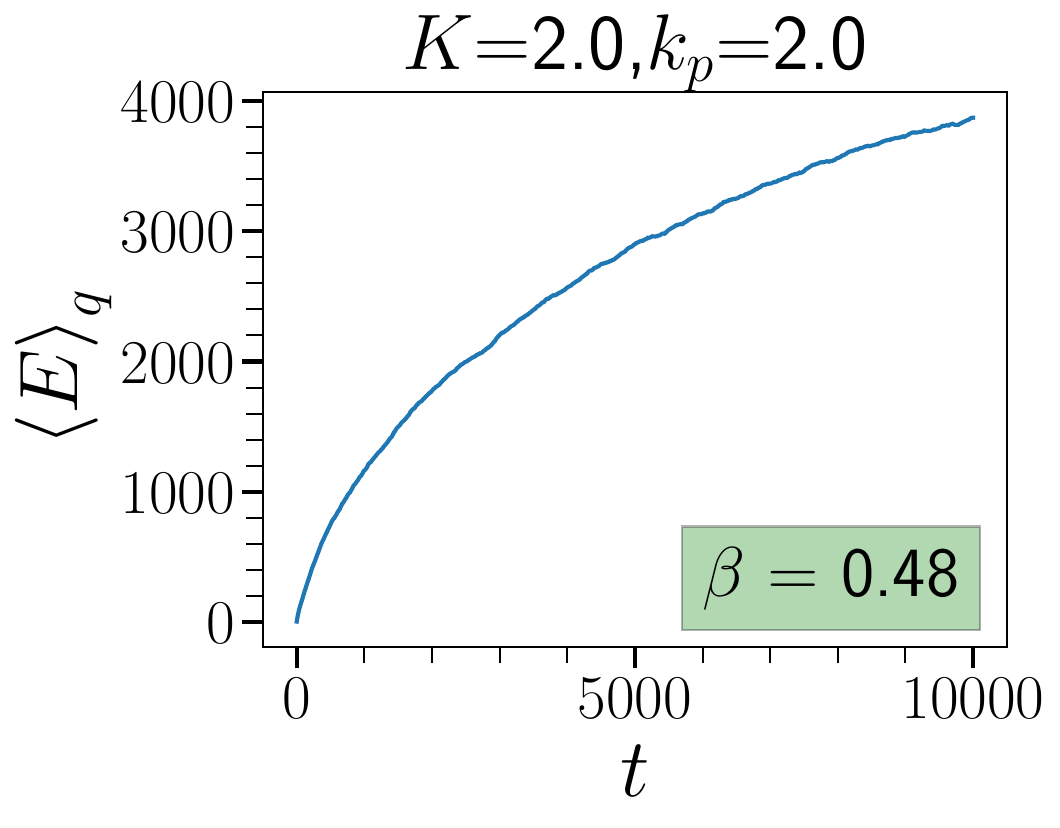}
    \includegraphics[width=0.46\linewidth]{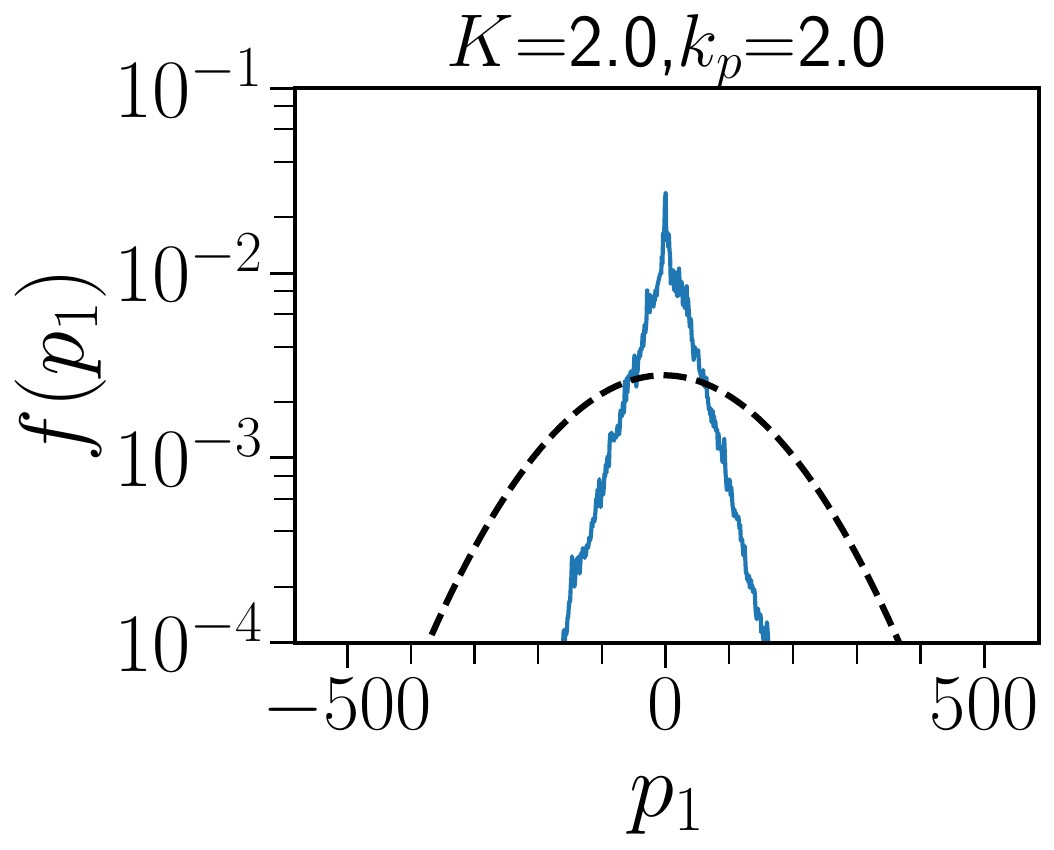}
    
	\includegraphics[width=0.46\linewidth]{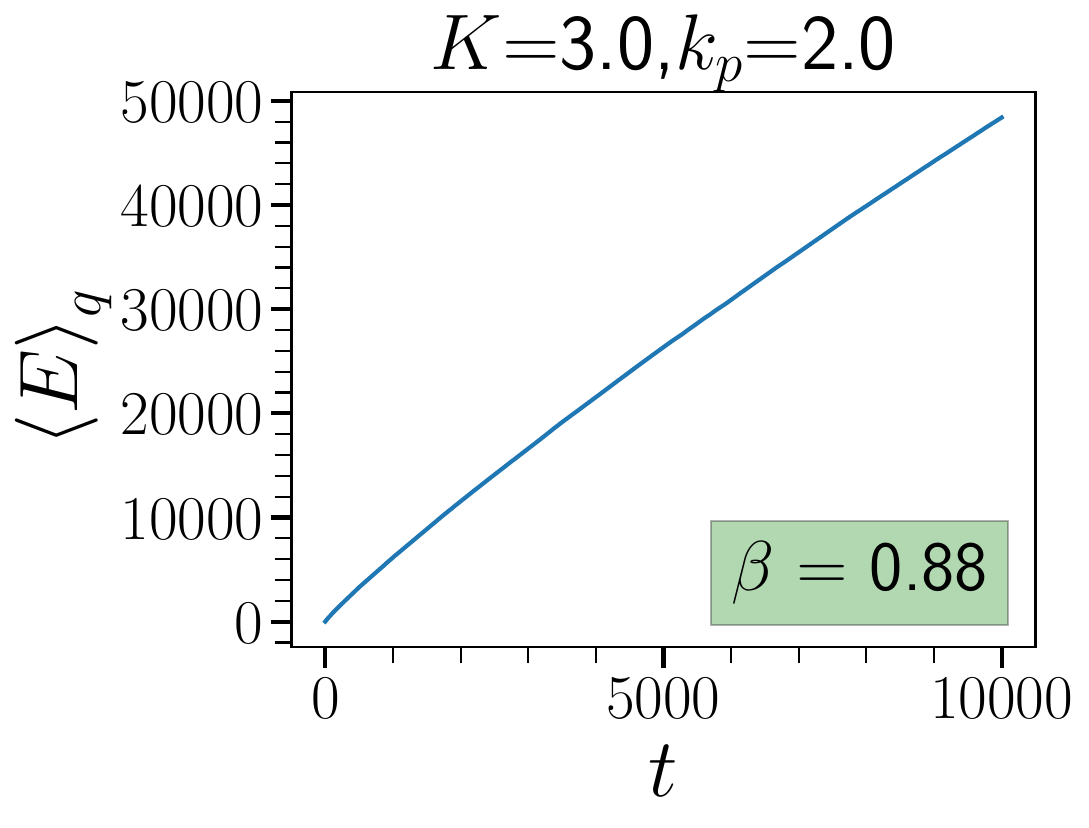}
    \includegraphics[width=0.46\linewidth]{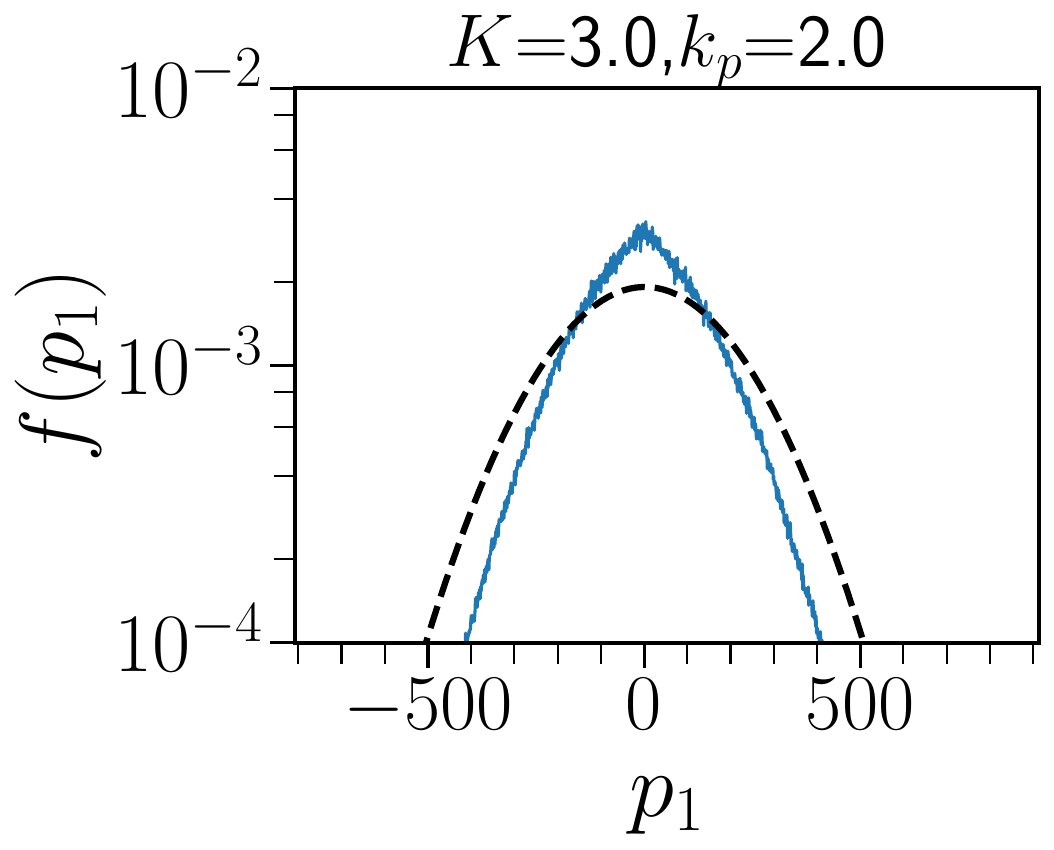}
    
	\includegraphics[width=0.46\linewidth]{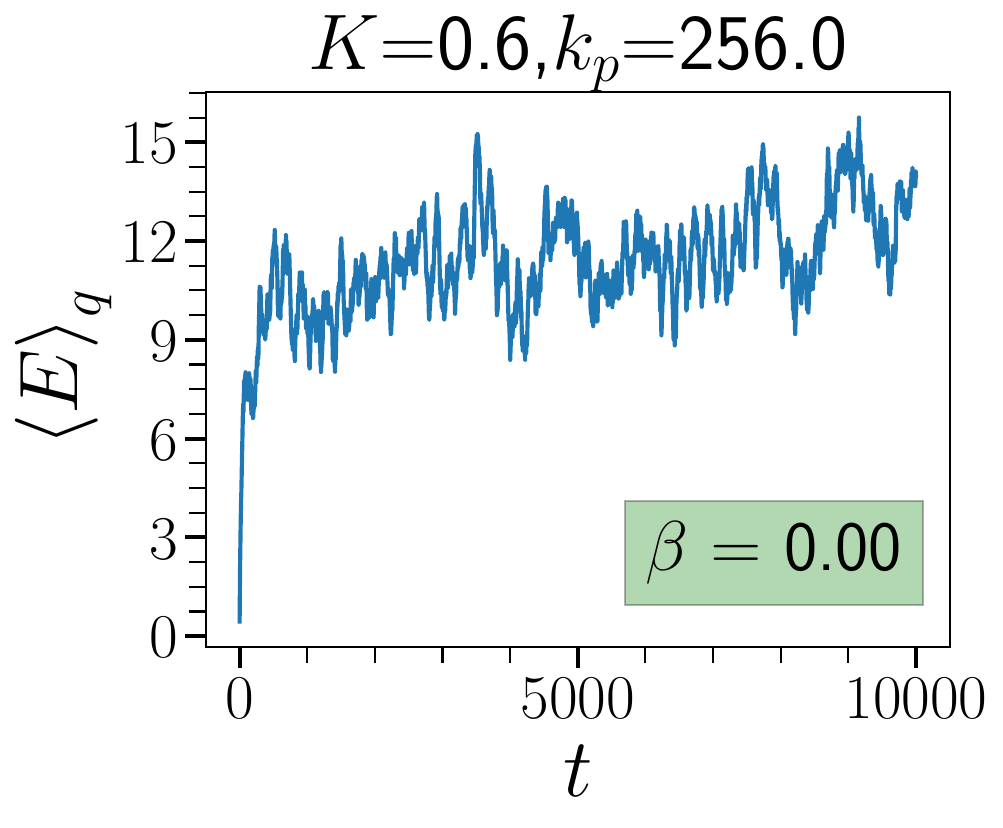}
    \includegraphics[width=0.46\linewidth]{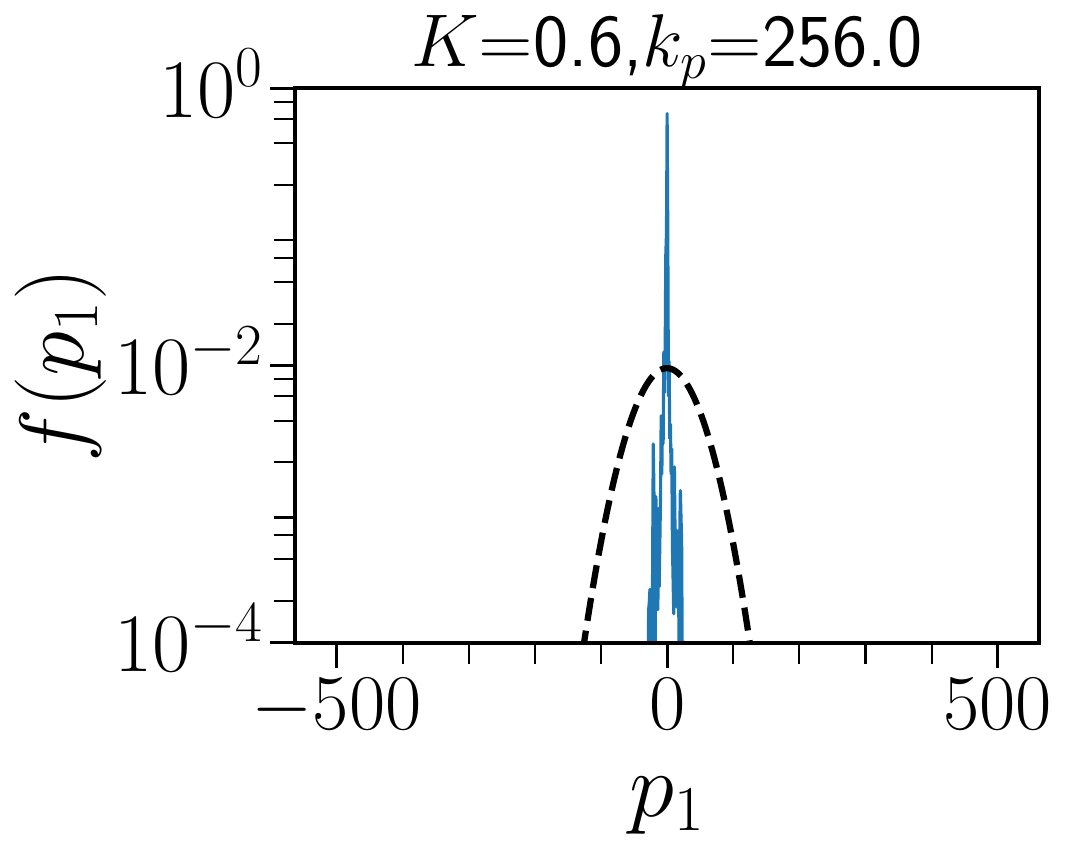}

    \begin{picture}(0,0)
	\put(-88,454){(a)}
    \put(34,455){(b)}
    \put(-89,362){(c)}
	\put(34,361){(d)}
	\put(-86,267){(e)}
    \put(34,270){(f)}
    \put(-84,173){(g)}
    \put(34,178){(h)}
    \put(-91,85){(i)}
    \put(34,82){(j)}
	\end{picture}
    
\caption{(left panel) The evolution of quantum mean energy $\langle E\rangle_{q}$. (right panel) shows the corresponding classical (black dashed line) and quantum momentum distribution (blue line) $f(p_1)$ after $10^{4}$ kicks. The parameters $(K, k_p)$ (shown on top of each figure) correspond to that shown in Fig. \ref{fig:1} for classical dynamics. Since the energy growth follows $\langle E\rangle_{q} \sim t^{\beta}$, the numerically estimated $\beta$ is also shown in left panel. The other parameters are $\alpha_{1}=\sqrt{3}$, $\alpha_{2}=\sqrt{5}, \hbar=1$.}
\label{fig:2}
\end{figure}

In the numerical simulations, the initial state $\ket{\Psi_0(\mathbf{p})}$ is taken to be a momentum eigenstate which, in the momentum basis, is simply a $\delta$-function placed at $(p_1, p_2) = (0,0)$. The state of the system after the $t$-th kick is given by $\ket{\Psi_t(\mathbf{p})} = \widehat{F}^{t} \ket{\Psi_0(\mathbf{p})}$. The Floquet operator is applied using a split-evolution technique \cite{Tan08} which requires an application of one Fourier and one inverse Fourier transform in each kick cycle. The evolution of quantum mean energy $\langle E\rangle_{q} $ is estimated as
\begin{equation}
    \big\langle E\big\rangle_{q}  = \big\langle \Psi_{t} \big| \widehat{p}_1^2 + \widehat{p}_2^2 \big|\Psi_{t} \big\rangle,
\end{equation}
whose time evolution is shown in Fig \ref{fig:2} for the same values of $K$ and $k_p$ used in Fig. \ref {fig:1}.

\section{Localization in $m$-LKR}
\label{sec:Loc_mLKR}
Now, we focus on the quantum evolution of an initially localized wavepacket. Figure \ref {fig:2}(a) shows a case for $K=0.2$ and $k_p=1$ for which the corresponding classical phase space (Fig. \ref {fig:1}(a)) is regular. The quantum mean energy does not increase with time ($\beta \approx 0.00$) but fluctuates about a mean value. This strongly correlates with the time-evolved wave function profile (Fig. \ref{fig:2}(b)), which remains strongly localized even after $10^4$ kicks had been applied. This type of localization is similar to that expected in the uncoupled (integrable) system with $k_p=0$ (or in spatially coupled LKR) where the quantum localization is strongly influenced by the absence of classical ergodicity, dominated by the presence of regular trajectories in the phase space. This is to be anticipated based on the semiclassical eigenfunction hypothesis \cite{Berry1977, Voros1979}, which posits that a wavefunction (in Wigner representation) would be pre-dominantly localized in regions visited by a typical classical trajectory for infinite times.

Figure \ref{fig:2}(c) shows the energy dynamics for $K=0.6$ and $k_p=2.0$. The quantum mean energy, though initially shown to increase with time, displays bounded oscillations. Asymptotically, on average, mean energy does not grow, {\it i.e.}, $\langle E\rangle_{q} \sim t^\beta$, with $\beta \approx 0.00$. As evident in Fig. \ref {fig:2}(d), the width of the quantum wavefunction profile is much smaller than for the corresponding classical momentum distribution. This must be compared with the diffusive classical mean energy shown in Fig. \ref{fig:1}(d). Despite the chaotic classical dynamics, the corresponding quantum system remains localized. This behaviour for the two-body $m$-LKR corresponds to the regime of dynamical localization. It is markedly different from the spatially coupled LKR since the observed dynamical localization is not just a manifestation of the underlying classical dynamics but rather a truly emergent quantum property similar to that observed in the single particle quantum kicked rotor \cite{SanPauBha22}.

In Fig. \ref{fig:2}(e),  for $K=k_p=2.0$, quantum dynamics is subdiffusive with $\beta \approx 0.48$. The corresponding wavefunction profile in Fig. \ref{fig:2}(f) reflects the fact that quantum diffusion is slower than the classical normal diffusion. For this choice of parameters, although dynamical localization is lost, a full-blown diffusive evolution is also absent. This is in contrast with the classical diffusive dynamics. If $k_p$ is kept constant, and $K$ increases further, quantum dynamics is delocalized and approaches near-linear quantum diffusion. This is the case illustrated in Fig. \ref{fig:2}(g,h) for $K=3$ and $k_p=2$. The energy growth is subdiffusive with $\beta=0.88$, and the wavefunction acquires a Gaussian profile (Fig. \ref{fig:2}(h)). For the same set of parameters, classical dynamics displays linear diffusion (Fig. \ref{fig:1}(f)).

Finally, Fig. \ref{fig:2}(i,j) shows for $k_p \gg 1$ a case that would classically correspond to the limit $K_s = K k_p \gg 1$ and therefore display classical chaotic diffusion. However, for sufficiently small $K$, we observe that the quantum energy growth saturates $(\beta \approx 0.00)$. The wave function displays an exponential profile (Fig. \ref{fig:2}(j)) whose width is far smaller compared to its classical counterpart. This is also quantum-induced two-body dynamical localization regime, similar to the case of $K=0.6$ and $k_p=2.0$ in Fig. \ref {fig:2}(c,d). It must be noted that dynamical localization can be sustained even if $k_p \gg 1$ as long as $K$ remains sufficiently small. This implies the existence of a threshold value of $K$ below which the quantum dynamics is always localized (notwithstanding the classical normal diffusion) for all values of $k_p$. Above this threshold, for larger $K$, localization is broken as $k_p$ increases and ultimately gives way to either quantum subdiffusion or diffusion.

Thus, in this interacting quantum system, as seen in Fig. \ref{fig:2}, we are able to observe rich dynamical features ranging from localization to linear diffusion upon varying parameters. In the next subsection, we discuss a global perspective of these features.

\begin{figure}[t!]
	\includegraphics[width=\linewidth]{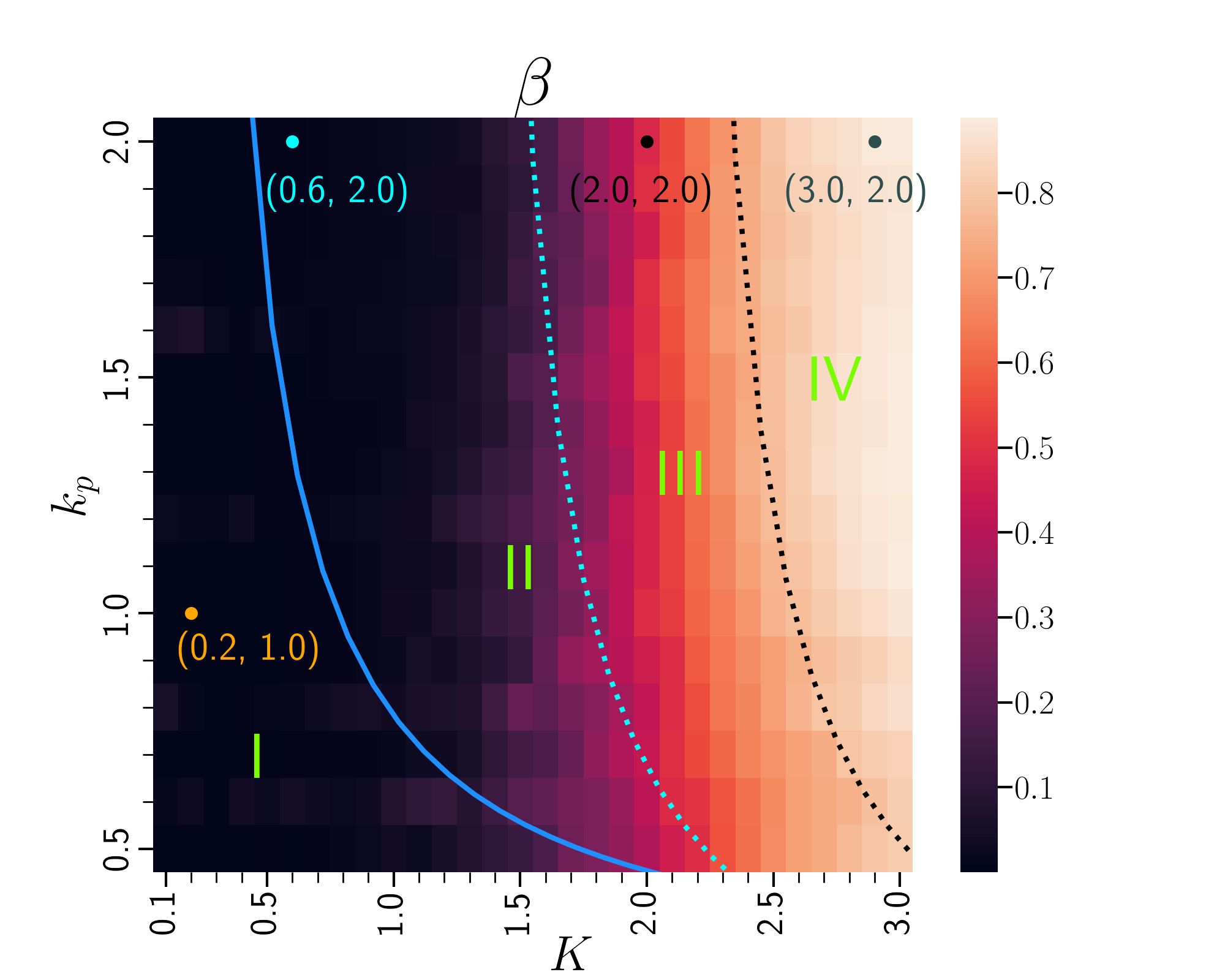}
\caption{Image map of exponent $\beta$ is plotted as a function of $k_{p}$ and $K$. The other parameters are $\alpha_{1}=\sqrt{3}, \alpha_{2}=\sqrt{5}$ and $\hbar=1$. This plot is obtained by evolving an initial wavepacket for $t=10^4$ kicks. The observed quantum mean energy growth is regressed against $\langle E \rangle_{q} \propto t^\beta$ to numerically estimate $\beta$. The four highlighted points correspond to the parameters chosen for illustration in Figs. \ref{fig:1} and \ref{fig:2}. Classical dynamics in region I is regular, and is chaotic in regions II to IV. See text for a detailed explanation of quantum dynamics in regions I to IV. The dashed lines demarcating regions II, III and IV are approximate indicators of boundaries.}
\label{fig:3}
\end{figure}

\subsection{Dynamical phases in parameter space}
For the quantum evolution, a global picture of the dynamical behaviours can be obtained from the image map displayed in Fig. \ref {fig:3}. Based on Eq. \ref {Eqn:transp_1}, the exponent $\beta$ corresponding to the growth of $\langle E \rangle_q$ is (numerically) determined as a function of $K$ and $k_{p}$ and is shown as a image map. Several broad quantum dynamical regimes are also marked on this image map. Note that $m$-LKR is integrable if $k_p=0$ or $K=0$. From Eq. \ref{largeev}, it is clear that classical chaos dominates for $K k_p > 1$. Thus, the (solid) blue line in Fig. \ref {fig:3} is the relation $K=1/k_p$. It demarcates classically chaotic and regular regimes, where region I is regular, while regions II to IV are chaotic. 

In Fig. \ref{fig:3}, region I correspond to classical regularity and $\beta \approx 0$. In this regime, the quantum localization results from wave packet evolution trapped within the confines of non-ergodic classical trajectories. This type of localization is realized in the limit of $K \to 0$ or $k_p \to 0$ both of which correspond to the semi classical limit. Hence, despite strongly localized wave packets, this should not be regarded as the regime of localization arising from purely quantum effects. 

\begin{figure}
    \centering
    \includegraphics[width=0.47\linewidth]{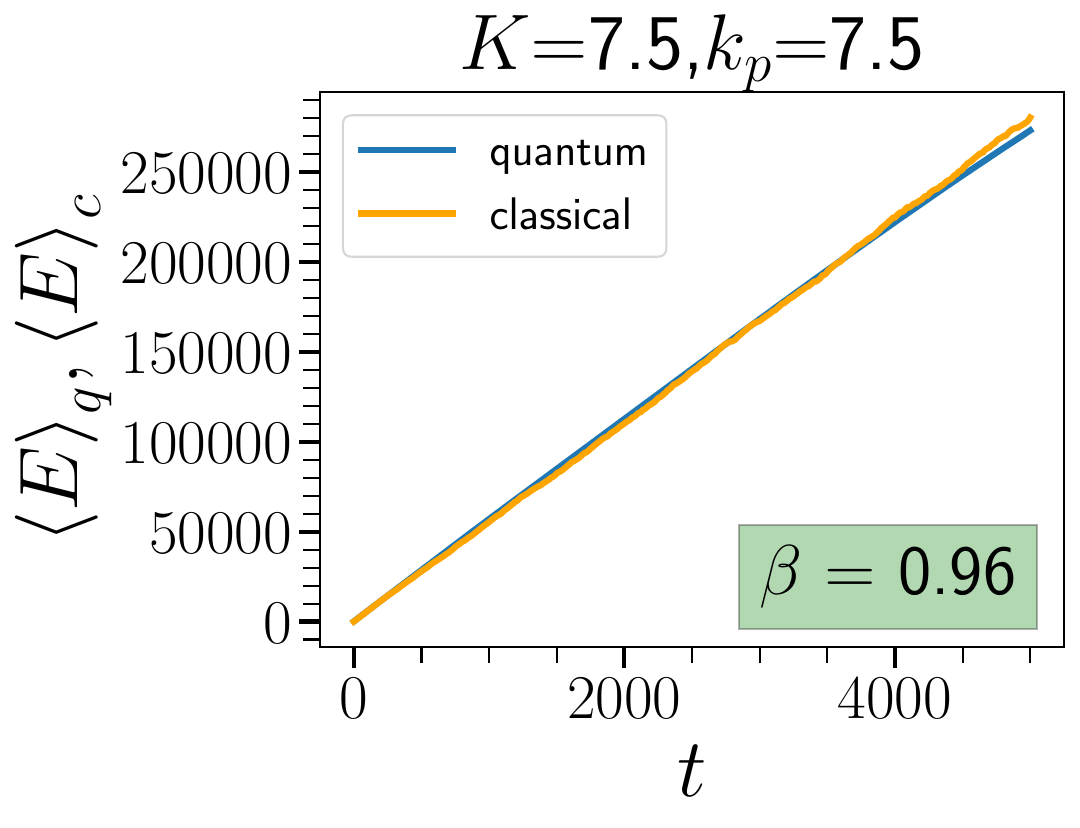}
    \includegraphics[width=0.47\linewidth]{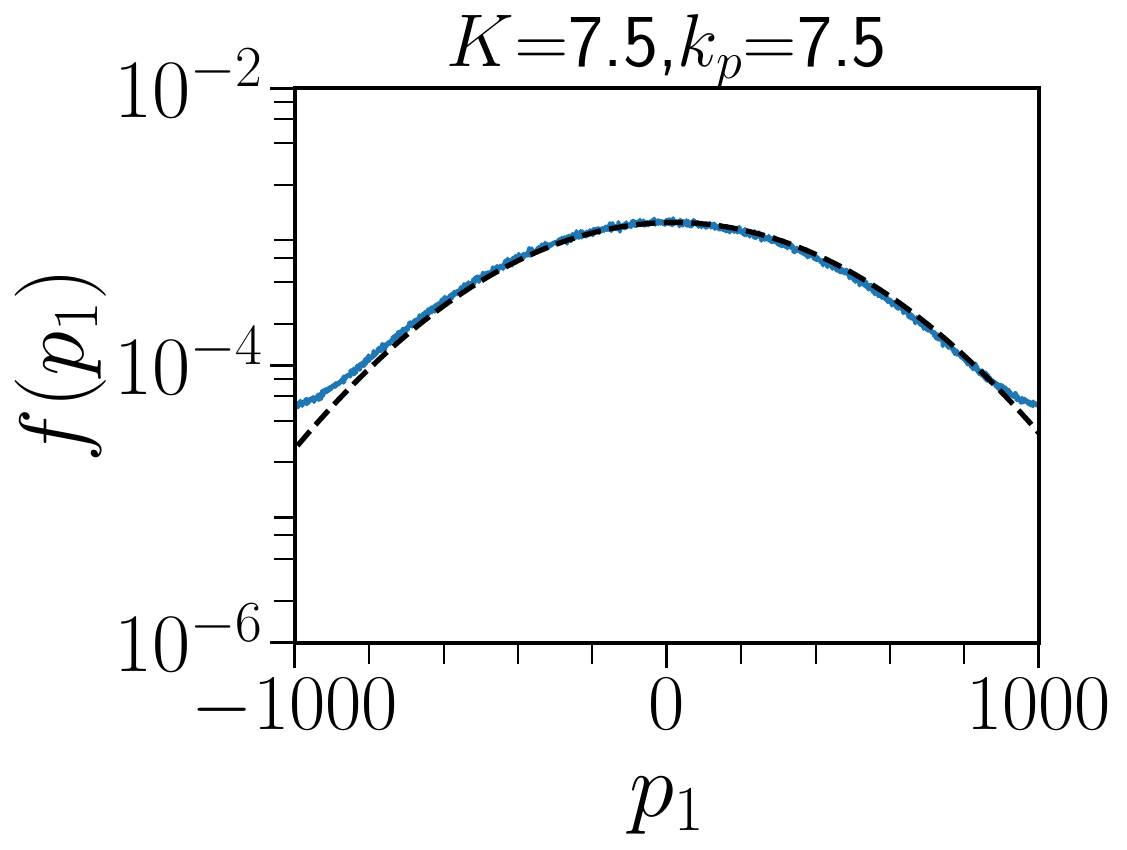}
    \begin{picture}(0,0)
	\put(-137,60){(a)}
    \put(-25,60){(b)}
	\end{picture}
    \caption{(a) Diffusive mean energy growth for classical and quantum model. Both display similar growth $\sim t^\beta$, with $\beta \approx 1.0$. (b) The classical (black dashed line) and quantum (blue line) momentum distributions (at $t=5000$) have a Gaussian profile. Other parameters are $\alpha_1=\sqrt{3}$, $\alpha_2=\sqrt{5}$, $\hbar=1$.}
    \label{fig:5}
\end{figure}

Regions II to IV fall in the classically chaotic regime. The quantum evolution in region II displays localization with $\beta \approx 0.00$. This is the two-body dynamical localization regime arising from purely quantum interference effects provided $k_p > 0$. The parameter choices shown in Figs. \ref{fig:2}(d,j) fall in regime II with $K, k_p > 0$. Their wavefunction profiles are exponential, while the corresponding classical momentum distributions are Gaussian. Indeed, as in the case of the quantum kicked rotor problem, in this regime quantum dynamics ignores the underlying classical features. Consistent with Fig. \ref{fig:3}, we find numerical indications that the two-body dynamical localized phase persists even for small values of the kicking strength $K$ provided $k_p \gg 1$. 

In contrast to localization regimes described above, in region III quantum wave-packet evolution is sub-diffusive with $\beta$ lying in the range 0.2 to 0.8. It might be recalled that recent experiments \cite{TohHuiMcc2022,CaoRosMas2022} in a different interacting system resulted in sub-diffusive dynamics rather than normal diffusion. This implies that localization is broken but, contrary to expectations, a full diffusive regime is not restored by interactions. Region III falls in this class because breaking integrability did not restore normal quantum diffusion. Finally, region IV is the regime of (nearly) quantum normal diffusion with $\beta = 0.8 - 1.0$. This energy growth correlates with similar qualitative dynamics in the classical regime (compare figures \ref {fig:2}(g) and \ref {fig:1}(g)). Physically, in a coupled system such as the $m$-LKR, quantum diffusive regime occurs due to one subsystem acting as a source of ''environment'' for the other in the chaotic regime \cite{PauBhaSan2022}. Thus, the ``noisy'' inputs from the other rotor breaks any localizing mechanism based on quantum coherence. If $K$ and $k_p$ are increased even further, we observe an almost exact correspondence between classical and quantum energy growth as evident from Fig. \ref {fig:5}. We emphasize that both these boundaries, between region II to III and III to IV, are not sharp, and the dashed lines are only indicative, and should not be regarded as representing any kind of critical transition. 



\section{Entanglement production in $m$-LKR}
\label{sec:EE}

How does entanglement evolve in $m$-LKR? In this section, we report the entanglement dynamics using von Neumann entropy to characterize localized and thermal phases in this two-body quantum system \cite{HorHorHor09,Laf16}.  Entanglement dynamics is well-studied in interacting systems with {\it finite} Hilbert space dimension, much less is known for the systems in infinite dimensional Hilbert spaces such as the two-body $m$-LKR. 
Main object of interest is the von Neumann entropy given by
\begin{equation}
    S(t)=-\operatorname{Tr}_{1}\left(\rho_{1}(t) \log \rho_{1}(t)\right)
\end{equation}
to characterize the entanglement between the subsystems, with each LKR representing a subsystem. In this, $\rho_{1}(t)=\operatorname{Tr}_{2} \rho(t)$ is the reduced density matrix obtained by tracing out the second subsystem, and $\rho(t)=|\Psi(t)\rangle\langle\Psi(t)|$ is the total density matrix of the time-evolved state $|\Psi(t)\rangle$. Let $N_1$ and $N_2$ represent subsystem dimensions such that $N_1 > N_2$ (a consequence of $\alpha_1 \ne \alpha_2$).
In the quantum chaotic regime, random matrix theory (RMT) estimates the average entanglement to be  $S_{\rm RMT} \approx \ln(\gamma(Q) N_1)$ \cite{Aru01}, where $\gamma(Q)$ depends on the ratio $Q = N_2/N_1$.

To compare the numerically estimated $S$ for $m$-LKR with RMT based prediction, an {\it effective} Hilbert space dimension $N^{\rm eff} = N_1^{\rm eff} N_2^{\rm eff}$ must be identified.  As the system evolves,  $N^{\rm eff}$ changes with time. Hence, the actual dimension (which is infinite) is replaced with $N^{\rm eff}(t)$ at time $t$. The effective dimension of $j$-th subsystem -- representing the occupation of states in Hilbert space indexed by $m$ -- is 
\begin{equation}
N_j^{\rm eff}(t) \approx \frac{1}{\sum_{p_{j}}(\sum_{p_{k}}|\Psi_{p_{1},p_{2}}(t)|^{2})^{2}} = \frac{1}{\sum_{p_{j}}(f(p_{j},t))^{2}} 
\label{eq:IPR}
\end{equation}
where $(j, k) = (1,2)$ or $(2,1)$, $\Psi_{p_{1},p_{2}}(t)$ is the time-evolved wavefunction, while $f(p_{j},t)$ denotes the corresponding marginal probability density.

With this information, the RMT average of entanglement can be expressed as
\begin{align}
S_{\rm RMT}(t) \approx \ln \left[ \gamma(Q) ~ N_1^{\rm eff}(t) \right],
\label{eq:entropy}
\end{align}  
where $\gamma(Q)$ is a time-dependent quantity. If the Hilbert space is finite, then $\gamma(Q)$ is a constant for given $N_1$ and $N_2$ \cite{BanLak04}; in particular, if $N_1=N_2$, then $\gamma \approx 0.6$.

\begin{figure}[h!]
    \includegraphics[width=0.8\linewidth]{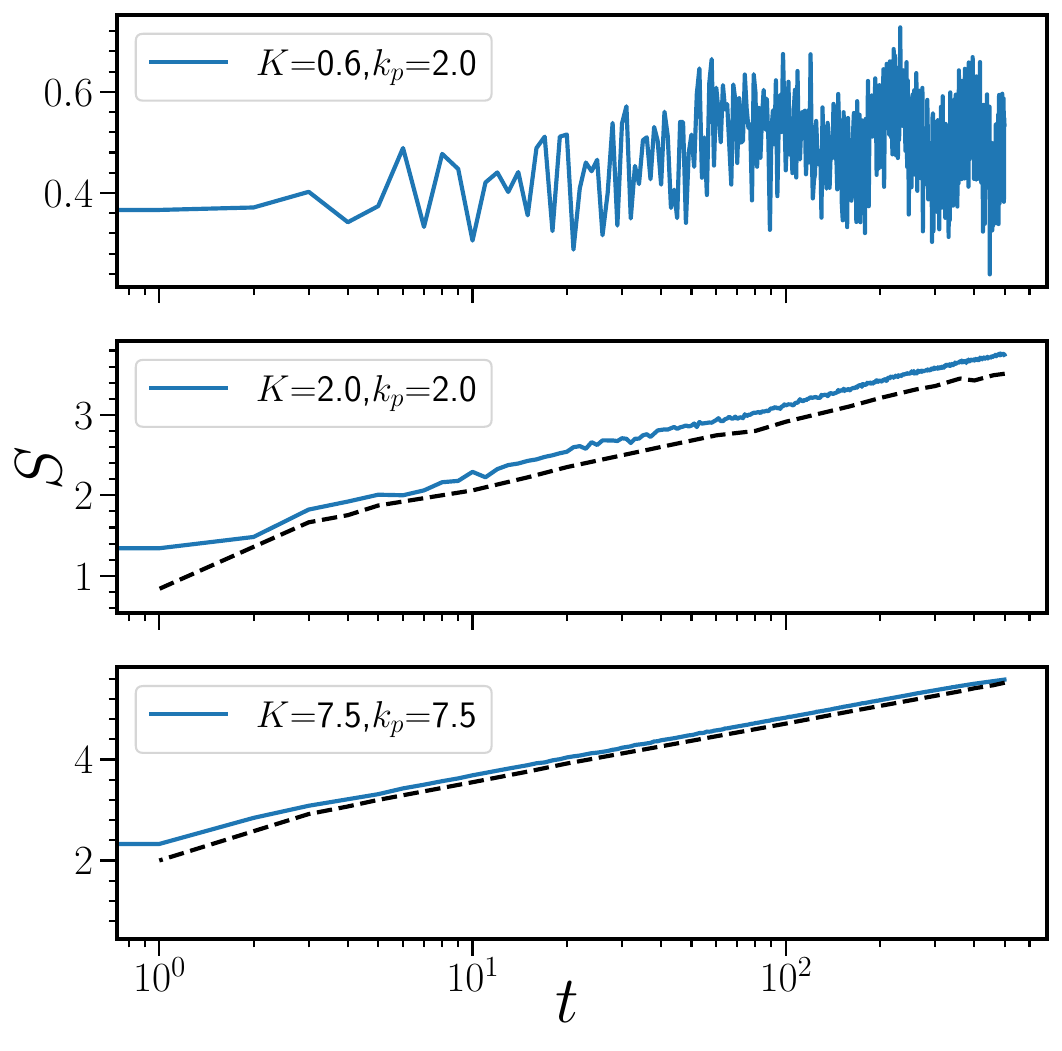}
       \begin{picture}(0,0)
	\put(-175,160){(a)}
	\put(-175,100){(b)}
	\put(-175,45){(c)}
	\end{picture}	
    \caption{Numerically computed entanglement $S(t)$ for $m$-LKR (solid blue line).
     (a) $(K, k_p) = (0.6, 2.0)$ corresponds to dynamical two-body localized phase and saturated entanglement, (b) (2.0,2.0) corresponds to quantum subdiffusive energy growth regime and logarithmic entanglement production, (c) (7.5,7.5) corresponds to quantum diffusive energy growth regime and logarithmic entanglement production. The black dashed line is an analytical estimate $S_{\rm RMT}(t)$ obtained by using Eq. \ref{eq:IPR} in Eq. \ref{eq:entropy}. The other parameters of $m$-LKR are $\alpha_{1}=\sqrt{3}, \alpha_{2}=\sqrt{5}$, $\hbar=1$.}
    \label{fig:6}
\end{figure}

Figure \ref{fig:6} displays entanglement entropy as a function of time for three cases. (a) For $K = 0.6, k_p=2$, a quantum steady state in the form of localization emerges, and entanglement saturates though it fluctuates about an average value. Entropy production is effectively stops. Consistent with Figs. \ref{fig:2}(c,d), this is another indicator of emergent dynamical localization. 
(b) For $K = 2, k_p=2$, in Figs. \ref{fig:2}(e,f), and for $t \gg 1$, the $S(t)$ has logarithmic dependence on time. The RMT estimate obtained using Eq. 15 and 16, shown as the dashed line, deviates from the numerically computed entanglement. However, it provides a reasonable estimate of the entanglement growth rate. 
Let us recall that $K = 2, k_p=2$ corresponds to quantum sub-diffusion, and the evolving wavefunction has an (approximately) exponential profile as seen in Fig. \ref{fig:2}(f). This cannot be regarded as a random state in the sense of RMT. This observation raises interesting questions and suggests that the RMT average in Eq. \ref{eq:entropy} is useful even when the system is not consistent with RMT-type assumptions.
(c) For $K=k_p=7.5$, the quantum diffusive behaviour manifests in logarithmic growth of entanglement (Fig. \ref{fig:6}(c)). As in the case of Fig. \ref{fig:6}(b), our RMT estimate is consistent with the numerically computed entanglement production rate (blue line). 

In both the sub-diffusive and diffusive cases (Figs. \ref{fig:6}(b,c)), logarithmic entanglement growth arises almost entirely from $N_1(t)$, while the ratio $Q=N_2(t)/N_1(t)$ is approximately a constant for our choice of parameters. However, the growth is faster for $K=k_p=7.5$ compared to that of  $K = 2, k_p=2$. This is not entirely surprising since the former corresponds to a larger Lyapunov exponent than the latter, and as noted in many earlier studies, this classical property leaves its footprint in the short-time asymptotics of entanglement as well \cite{WanGhoSanHu04}. This particular regime can be linked to classical behavior, where one of the rotors functions as a sort of noisy environment for the other \cite{SanArn20}.


The asymptotic entanglement saturation is well-understood in {\it finite} quantum systems. Figure \ref{fig:6} shows that, in unbounded Hilbert spaces, entanglement growth is linear only for very short timescales, and does not saturate as $t \to \infty$ (Fig. \ref{fig:6}(b-c)), except when a localized phase emerges (Fig. \ref{fig:6}(a)). In the former cases, absence of entanglement saturation is a direct consequence of the unbounded Hilbert space.
Remarkably, even if the entanglement growth rate does not vanish, our work demonstrates that its instantaneous growth rate coincides with that of $S_{\rm RMT}(t)$ estimated using effective Hilbert space dimension in Eq. \ref{eq:IPR}. This highlights the distinct behavior of entanglement dynamics in infinite dimensional Hilbert spaces from that of finite quantum systems.

\section{Conclusions}
\label{sec:conclusions}
The main motivation behind this work is to induce nonintegrability and chaos in interacting linear kicked rotor model and study the quantized version to explore the emergence and sustainability of dynamical localization and other phases. We address this question by studying an interacting two-body linear kicked rotor, and hence our results strictly pertain to this two-body system. Recently, it was shown that many-body linear kicked rotors interacting through spatial variables, under some mild conditions on the parameters, would always have integrals of motion and would not display chaotic regimes. In the quantum regime, this system displays many-body localization arising from the presence of integrals of motion. In order to probe localization in the chaotic limit, firstly, we create a version of a two-body linear kicked rotor that can display chaos. Since chaos is ruled out if spatial variables are coupled, we show that the momentum-coupled two-body linear kicked rotor ($m$-LKR) becomes chaotic. We confirm this by analytically estimating the largest Lyapunov exponent $\lambda_{\rm max}$, which turns out to be in good agreement with that computed through simulations. In particular, $\lambda_{\rm max} \approx \ln(K_s)$, where $K_s$ is the scaled chaos parameter. Further, the nature of classical dynamics (regular or chaotic) is also consistent with the temporal evolution of classical mean energy as a function of time. In the chaotic regime, mean energy growth is diffusive in nature.

The quantum dynamics of $m$-LKR provide evidence for the existence of four different regimes, including two-body localized regimes. We briefly summarize them here : (1) If the parameters are chosen to be in the near-integrable regime $(K_s < 1)$, then localization-like behaviour is observed, arising due to the strong influence of the regular classical dynamical features. This is effectively a semi-classical localization effect and is not a purely quantum effect. For values of $K_s$ greater than unity, while the classical dynamics is chaotic, the quantum dynamics displays three different dynamical regimes. (2) First is the regime of two-body dynamical localization in which the wave function has an exponential profile. For the quantum mean energy, $\langle E\rangle_{q} \sim t^\beta$, here $\beta = 0$ showing dynamical localization. This regime persists for interaction strength $k_p \gg 1$ where $K_s = Kk_p \gg 1$, and so is in the classically chaotic regime displaying classical linear diffusion, so long as kicking strength $K$ is sufficiently small. (3) Beyond these two types of localized phases, we observe two other parametric regimes in which classical dynamics is chaotic, but the quantum localization is not sustained. Firstly, one of them corresponds to quantum sub-diffusion with mean energy $\langle E\rangle_{q} \sim t^\beta$, where $0 < \beta < 1$, though the classical energy growth is diffusive. (4) Finally, quantum diffusive regime occurs with $\langle E\rangle_{q} \sim t^\beta$ and $\beta=1$. In this case, the classical energy growth is also diffusive. Quantum energy growth closely follows classical energy growth.

We have also shown that the instantaneous entanglement production rate can be estimated using random matrix average and an effective Hilbert space dimension at any arbitrary time. In the localized phases, the entanglement production saturates, while in the quantum (sub-)diffusive phases the entropy production has a logarithmic dependence on time. Thus, in summary, the modified momentum coupled model introduced in this work displays a variety of transport properties ranging from localization and quantum sub-diffusion to diffusion, depending on the choice of parameters. While we have demonstrated this for ease and convenience of visualization (especially the Poincar\'e maps) using a two-body system, we expect these features to exist in the many-body version of the linear kicked rotor as well. 

The results presented here will have significant implications for studies being carried out on many-body systems for which it is not straightforward to explore their classical limit. Popular examples of such systems would be the interacting spin chains. In such models, It is important to distinguish between dynamical localization and localization-like effects induced by underlying classical dynamical structures. Further, even in the presence of nearly complete chaos, interactions can sustain dynamical localization, or they can display quantum diffusion. In a future study, it would be interesting to explore the general conditions under which each of these regimes can be realized. If an experimental realization \cite{KesGanRef16} of $m$-LKR becomes possible, then it would lead to the verification of transport and localization features reported in this paper. \\

\acknowledgments{MSS acknowledges the support of MATRICS Grant MTR/2019/001111 from Science and Engineering Research Board (SERB), DST, Government of India. All the authors acknowledge the National Supercomputing Mission for the use of PARAM-Brahma at IISER Pune.  JBK would like to thank Sanku Paul, Sreeram PG and Harshini Tekur for helpful discussions.}

\bibliography{ref}

\end{document}